\documentstyle[multicol,aps,prl,floats]{revtex}

\draft

\begin{document}
\draft
\title{{\tenrm\hfill Submitted to Phys. Rev. E}\\
Inverse cascade and intermittency of passive scalar in 1d smooth flow}
\author{M. Chertkov$^a$, I. Kolokolov$^b$ and M. Vergassola$^{c}$}
\address{$^{a}$ Physics Department, Princeton University, Princeton, 
NJ 08544, USA.\\
$^{b}$ Budker Institute of Nuclear Physics, Novosibirsk 630090, Russia.\\
$^{c}$ CNRS, Observatoire de Nice, B.P. 4229, 06304 Nice Cedex 4, France.}
\date{June 12, 1997}
\maketitle

\begin{abstract}
Random advection of Lagrangian tracer scalar field $\theta (t,x)$ by a
one-dimensional, spatially smooth and short-correlated in time velocity
field is considered. Scalar fluctuations are maintained by a source
concentrated at the integral scale $L$. The statistical properties of both
scalar differences and the dissipation field are analytically determined,
exploiting the dynamical formulation of the model. The Gaussianity known to
be present at small scales for incompressible velocity fields emerges here
at large scales ($x\gg L$). These scales are shown to be excited by an
inverse cascade of $\theta ^{2}$ and the probability distribution function
(PDF) of the corresponding scalar differences to approach the Gaussian form,
as larger and larger scales are considered. Small scales ($x\ll L$)
statistics is shown to be strongly non-Gaussian. Collapse of scaling
exponents for scalar structure functions\thinspace takes place: moments of
order $p\ge 1$ scale all linearly, independently of the order $p$. Smooth
scaling $x^{p}$ is found for $-1<p<1$. Tails of scalar differences PDF are
exponential while, at the center, a cusped shape tends to develop when
smaller and smaller ratios $x/L$ are considered. The same tendency is
present for scalar gradients PDF with respect to the inverse of the
P\'{e}clet number (the pumping-to-diffusion scale ratio). The tails of the
latter PDF are however much more extended, decaying as a stretched
exponential of exponent $2/3$, smaller than unity. This slower decay is
physically associated with the strong fluctuations of the dynamical
dissipative scale.
\end{abstract}


\renewcommand{\thesection}{\arabic{section}}

\section{Introduction}

\label{s:introduction}

Small-scale statistics of a passive scalar advected by a large-scale
incompressible velocity field is an old problem in turbulence theory \cite
{59Bat,68Kr}. Scalar fluctuations are maintained by a large-scale forcing,
with typical scale $L$. According to the classical picture \cite{49Obu} of
direct cascade of the scalar, the injected scalar is mainly transferred
downscales to the convective interval and then to the dissipative region.
For smooth velocities, the statistics of this scalar transfer can be
analyzed systematically and has been characterized in much detail in \cite
{68Kr,94SS,95CFKLa}. The core of the one-point scalar probability
distribution functions (PDF) is Gaussian with variance $O(\log {\rm Pe})$,
where the P\'{e}clet number ${\rm Pe}$ is very large \cite{95CFKLa}. Far
tails of the PDF decay exponentially \cite{94SS,95CFKLa}. The physical key
ingredient at the basis of these results is that material lines are
stretched, i.e. the maximum Lyapunov exponent $\overline{\lambda }$ for
particles separation is positive. Typical trajectories will therefore be
exponentially stretched and dynamically contracted trajectories are so rare,
that they can only affect the extreme tails of the statistics. On the other
hand, much interest has been recently attracted by Kraichnan model \cite
{94Kr} for its intermittent scaling behavior \cite{95CFKLb,95GK,95SS}. The
picture emerging there is that dynamically contracted Lagrangian
trajectories play a crucial role for structure functions scaling exponents $%
\zeta _{p}$ and, thus, for intermittency. The constant asymptotic behavior
of $\zeta _{p}$ for large orders $p$ found in \cite{97Che} comes, for
example, from the most contracting possible trajectories and the value $%
\zeta _{\infty }$ from nontrivial fluctuations of the degree of freedom
constrained by incompressibility to still be dynamically stretched.

Previous remarks have led us to investigate scalar transport in smooth
compressible flow. The motivation is that compressibility might slow down
Lagrangian separations and, thus, lead to nontrivial scaling and
intermittency properties. These can be then analyzed systematically, using
techniques specific for smooth velocities. Positive Lyapunov exponents are
indeed characteristic of isotropic, solenoidal flow \cite{Cocke,SAO}, but
this property might be lost when compressible flow are considered. It is,
for example, known that for compressible flow a substantial slow-down of
large-times transport can take place (see \cite{VA}). Since trapping effects
are amplified when the dimensionality of space is low, we have focused our
attention on the one-dimensional case. More specifically, we have considered
the smooth limit of the one-dimensional $\delta $-correlated in time model
recently introduced in \cite{97VM}. In the absence of pumping and
dissipation, any function of the tracer $\theta (t,x)$ (say temperature) is
advected along Lagrangian trajectories and globally conserved in average
(provided the velocity is temporarily fast or spatially smooth). Switching on
the energy ($\theta ^{2}$) supply at the integral scale, one expects that a
steady (or quasi-steady, as discussed in Section~\ref{ss:model})
distribution of the scalar is attained. The main question raised here is how
trapping effects due to compressibility affect the redistribution of energy
among the scales and the intermittency properties of the scalar field at the
stationary state. To answer this question, we exploit the dynamical
formulation of the model to calculate the statistical properties of $\theta $
both up and downscales, i.e. at the scales smaller and larger than the
integral scale $L$. Since the scalar is a tracer in the velocity field and
the velocity is smooth, the problem reduces to studying Lagrangian
separations statistics.

The major physical difference appearing with respect to the incompressible
case is that the maximum Lyapunov exponent for Lagrangian separations is
negative. This means that, along typical trajectories, distances are mostly
contracted and the stretching process is strongly intermittent in time. As
in \cite{85ZMRS}, the second and higher powers of the distance $R(t)$
between Lagrangian trajectories grow exponentially, while its low-order
positive moments decay exponentially. This is the dynamical origin of the 
major
results found in this paper\thinspace : inverse cascade and Gaussianity at
large scales and extreme intermittency at small scales.

Scalar correlations are indeed essentially governed by the statistics of the
time spent by particles at distances smaller than $L$. Consider then two
particles initially separated by a distance $x\gg L$. Despite of the fact
that $x\gg L$ initially, the distance $R(t)$ will typically reduce to $O(L)$
by a large time $\sim\log \left( x/L\right) /|\overline{\lambda }|$. The
consequence is that even scales much larger than $L$ are strongly excited
and this is the dynamical hint of the inverse energy cascade. Moreover, for
moments of order $n\ll \log \left( x/L\right)$, relative fluctuations around
the previous typical time are small and this leads to the Gaussianity, e.g.
of the core of scalar differences PDF. On the contrary, the statistics at
small scales $x\ll L$ is associated with the stretching process. The time to
reach the integral scale $L$ strongly fluctuates and this is reflected in
the intermittency of both scalar differences and gradients.

Note that the inverse cascade of the scalar taking place here differs in one
important respect from other known examples of inverse cascades (say, energy
cascade in 2d Navier-Stokes turbulence \cite{67Kr} and number of waves
cascade in wave turbulence \cite{92ZLF})\thinspace : no flux of another
integral of motion (like enstrophy or wave energy) is present. The origin of
the inverse cascade found here is purely dynamical and associated with
trapping effects. An interesting consequence of the inverse cascade is that
the equation for velocity differences PDF, derived here exactly from the
dynamics, coincides with the one without dissipative anomaly (operator
product expansion, which may result in the anomaly, was proposed in \cite
{95Pol} in the context of the Burgers turbulence; see also \cite{97Yaka} for
possible extensions to the passive scalar turbulence).
The absence of anomaly is indeed a consequence of the inverse cascade,
preventing the rare trajectories emerging from the dissipative range to
affect the convective range behavior.

Strong downscales intermittency emerges all over the quantities calculated
in Sections IV-VI. Moments of scalar differences of order $n\ge 1$ all scale
linearly with $x$, independently of the order. This collapse of exponents
carries over to the dissipation field $\epsilon $, that has all its integer
moments scaling with the same power of the P\'{e}clet number. Smooth scaling
is observed for low-order moments of both scalar differences and
dissipation. Very large fluctuations of these two quantities behave,
however, quite differently. Scalar differences PDF has indeed a Lorentzian
shape for values smaller than unity and exponential tails. The tails of the
dissipation field PDF are, on the contrary, stretched exponentials with
exponent $1/3$ (and not $1/2$). The additional probability for these strong
events comes from fluctuations of the dynamical dissipative scale.
Comparison of the $n$-th moment of the dissipation field with the $2n$-th
moment of scalar differences establishes indeed the effective viscous scale.
This appears to be a strongly fluctuating quantity, growing factorially with 
$n$ \footnote{%
An essential enhancement of the dissipative scale was observed also in \cite
{96FKLM}, where the tail of the scalar pdf was studied for the
incompressible case by an instanton technique.}. This factorial dependence
is the cause of the $1/3$ stretched exponential.

Even though the explicit calculations are
quite lengthy, the underlying technical ideas, that make the analytical
calculations doable, are simple to explain. Since the scalar is passive, 
its $n$-th order correlation function can be presented in terms of a matrix
element of an auxiliary ``Quantum'' mechanics. There are then 
two import steps in the evaluation of such matrix elements 
(and thus of the scalar PDF's).
First, smoothness of the velocity field (stretchings and compressions are
uniform on all the
particles) allows to reduce (Section \ref{s:lagrangian} and
Appendix \ref{s:DynForm}) the multi-particle to a one-parameter
problem. The fact that the Lyapunov exponent is negative clearly emerges
in this procedure. This leads, after the very direct calculations 
of Sections \ref{s:gradients} and \ref{s:differences}, to a compact 
expression for the PDF of the scalar differences in the convective interval.
Here, the temporal
dynamics of the fluctuating degree of freedom is local, while the 
locality is lost in the dissipative range. The second step comes then into
play. To describe the
convection-diffusion interplay in Section \ref{s:dissipation} and Appendix 
\ref{s:Longtime}, we use a scale separation procedure. The temporarily
nonlocal (diffusive) and local (convective) dynamics of the fluctuating
degree of freedom are well separated by time $t_{0},$ $1\ll t_{0}\ll \ln [%
\mbox{Pe}]$, if the P\'{e}clet number $\mbox{Pe}$ is large. Independence of
the resulted average (say, the gradient's PDF) over both local and nonlocal
domains on $t_{0}$, and smallness of the neglected terms with respect to the
inverse P\'{e}clet number, justify the scale separation procedure.

The plan of the paper is as follows. In the following section, the
one-dimensional passive scalar model is briefly recalled, its relevant time
and length-scales are discussed, and the inverse cascade issue is explained
from consideration of scalar pair correlation function. The dynamical
formulation associated with the passive scalar equation is the subject of
Section~\ref{s:lagrangian} and the Appendix \ref{s:DynForm}. The latter is
based on the Martin-Siggia-Rose formalism, while the former is in terms of
particles formalism. A key point arising in both procedures is that
compressibility couples the dynamics with a global mode. This mode must then
be taken into account in order to get the dynamical formulations.
Multi-point correlation functions of scalar gradients are discussed in
Section~\ref{s:gradients}. Scalar differences PDF is described in Section~%
\ref{s:differences}, where behaviors up and downscale with respect to $L$
are considered in two different Subsections. To describe the
advection-diffusion interplay, we develop a scale separation procedure in
Section~\ref{s:dissipation}, which is 
devoted to the PDF of dissipation. We use the
scale-separation procedure also in Appendix \ref{s:Longtime} to study the
question, how does the steady regime for the pair correlation function of
the gradients (discussed in Section \ref{ss:model}) form\,? The final
section is reserved for conclusions and discussion of questions which may be
of a general relevance for other problems in turbulence and physics of
disordered systems.

\section{The model. Pair correlation function and inverse cascade.}

\label{ss:model}

Our aim here will be, first, to briefly recall the equations of the
one-dimensional model introduced in \cite{97VM}, and then, solving the
equation for scalar pair correlation function, show the effects of
compressibility on the redistribution of energy among the scales and how
does the inverse cascade of the scalar work.

The advection-diffusion equation governing the evolution of the passive
scalar $\theta (t,x)$ is \footnote{%
Note that there are two types of passive fields for compressible flow.
Lagrangian tracers, like entropy or temperature (provided pressure is slowly
varying in space and time), are conserved along Lagrangian trajectories in
the absence of diffusion and pumping. The local maxima of the field do not
grow in the absence of pumping. Concentration fields, e.g. of a pollutant,
are, on the contrary, only globally conserved and their maxima can be
amplified. The equations for the two types of fields differ by the position
of the space-derivative in the velocity term. In our one-dimensional case,
the two possibilities correspond to the $\theta $ and the $\omega $ fields,
respectively.} 
\begin{equation}
\partial _{t}\theta +u\partial _{x}\,\theta =\kappa \partial _{x}^{2}\theta
+f.  \label{original}
\end{equation}
The velocity field $u$ and the force $f$ are both assumed to be Gaussian and 
$\delta $-correlated in time. The force has correlation function 
\begin{equation}
\langle f(t,x)\,f(t^{\prime },x^{\prime })\rangle =\delta (t-t^{\prime
})\chi \left( {\frac{x-x^{\prime }}{L}}\right) ,  \label{forza}
\end{equation}
spatially concentrated at the integral scale $L$. The velocity has zero
average and correlation function 
\begin{equation}
\langle u(t,x)u(t^{\prime },x^{\prime })\rangle =\left[ V_{0}-S(x-x^{\prime
})\right] \delta (t-t^{\prime })\qquad {\rm with}\quad S(x)=|x|^{2-\gamma }.
\label{corr}
\end{equation}
The specific smooth case considered here corresponds to $\gamma =0$. The
scaling behavior of the structure function $S$ persists up to the infrared
cut-off $L_{u}$, the largest scale in our problem. The scale independent
part of the velocity correlation function $V_{0}$ is estimated by the
infrared cutoff of the velocity field squared, $\sim L_{u}^{2}$. For scales
larger than $L_{u}$, velocity correlations decay to zero, i.e. the structure
function saturates to the constant value $V_{0}$. The equation of motion for
the gradient field $\omega =\partial _{x}\theta $ immediately follows from (%
\ref{original}) 
\begin{equation}
\partial _{t}\omega +\partial _{x}\left( u\,\omega \right) =\kappa \partial
_{x}^{2}\omega +\partial _{x}f.  \label{omega}
\end{equation}
The $\delta $-correlation in time of both the velocity and the forcing
allows to derive closed equations of motion for equal-time correlation
functions \cite{68Kr}. It is for example easy, e.g. using Gaussian
integration by parts (see \cite{95Fri}), to derive the equation for the
second-order correlations $\Omega (t,x)=\langle \omega (t,x)\omega
(t,0)\rangle $ and $F(t,x)=\langle \theta (t,x)\theta (t,0)\rangle $ 
\begin{eqnarray}
\partial _{t}\Omega &=&\partial _{x}^{2}\left[ \left( S(x)+2\kappa \right)
\Omega \right] -\partial _{x}^{2}\chi ,  \label{boh} \\
\partial _{t}F &=&\left( S(x)+2\kappa \right) \partial _{x}^{2}F+\chi .
\label{boh2}
\end{eqnarray}
It is convenient to consider first equation (\ref{boh}) and then recover the
correlations of the scalar by integration. From the very definition of the
correlation function, it follows that the solution of (\ref{boh}) is even in 
$x$. Looking for a stationary solution, there is another boundary condition
needed to fix the integration constants. This comes from the dynamics. Let
us indeed consider the situation when the system starts from rest ($\theta
\equiv 0$). At any subsequent time, the solution satisfies $\int \Omega
\,dx=0$, where the integral is taken over all the space. Note that such a
condition is consistent and actually dictated by the dynamics\thinspace :
the integral of the correlation function of the gradients is conserved in
the evolution on account of the double space derivative in the r.h.s. of (%
\ref{boh}). It is then easy to find the stationary solution of (\ref{boh}) 
\begin{equation}
\Omega (x)={\frac{\chi (x)}{2\kappa +S(x)}}-{\frac{C}{2\kappa +S(x)}}\qquad 
{\rm where}\quad C={\frac{\int \chi /(2\kappa +S)}{\int 1/(2\kappa +S)}},
\label{solution}
\end{equation}
(see Appendix \ref{s:Longtime} for a dynamical derivation of (\ref{solution})).
The expression (\ref{solution}) is very illuminating 
for several reasons. Let us first
consider the case where no infrared cutoff $L_{u}$ is present. The integral $%
\int 1/(2\kappa +S)$ is then convergent and, for small molecular
diffusivity, varies as $1/\sqrt{\kappa }$. For the sake of concreteness, let
us specifically consider here a forcing that is regular at the origin. The
dominant terms for distances much smaller than the integral scale $L$ are
then 
\begin{equation}
\Omega (x)={\frac{|\chi ^{\prime \prime }(0)|}{2(2\kappa +S)}}\left[ aL\sqrt{%
\kappa }-x^{2}\right] ,  \label{small}
\end{equation}
where $a$ is constant $O(1)$, dependent on the detailed form of the pumping.
The first and the second term dominate respectively at scales smaller and
larger than $L/\sqrt{{\rm Pe}}$, where the P\'{e}clet number ${\rm Pe}\equiv
L/\sqrt{2\kappa }$ is supposed to be very large. Note that this scale is
still much larger than the dissipative scale $L/{\rm Pe}$. The most
interesting aspect of (\ref{small}) is that the dissipation $\kappa \,\Omega
(0)$ vanishes as $1/{\rm Pe}$, i.e. there is no direct cascade. The energy%
\footnote{%
It is important to mention that, in the absence of pumping and dissipation,
the average over the velocity of any function of the scalar $\langle
f(\theta ) \rangle $ is conserved, although the average over all the space
of $f(\theta )$ itself is not conserved in the particular realizations.} is
actually transferred upscales by an inverse cascade, as also emerges from
the behavior of scalar correlations. Let us indeed insert into (\ref{boh2})
the correlation $\langle \theta (t,x_{1})\theta (t,x_{2})\rangle
=\int_{-\infty }^{x_{1}}\int_{-\infty }^{x_{2}}\Omega (y_{1}-y_{2})$ with
the $\omega $-correlation function having the expression (\ref{solution}).
It is easily checked that the energy satisfies $\partial _{t}\langle \theta
^{2}\rangle =C$, i.e. it grows linearly with time. Note that the growing
mode is constant in space and thus disappears when differences or gradients
are considered. The effect of the advective term $u\partial _{x}\theta $ is
therefore to transfer energy upwards in the scales. Since dissipation is
quadratic in the wavenumber, the energy on the large scales cannot
practically be dissipated and it is piled-up. This is at the origin of the
linear growth in time of $\langle \theta ^{2}\rangle $. $C\neq 0$ correponds
to the inverse cascade, which therefore holds generically for any scaling
exponent $\gamma <1$ in (\ref{corr}) (if the source function $\chi $ is not
exceptional).

Let us now introduce an infrared cutoff $L_{u}$ in the velocity field. Since 
$S$ saturates to a constant, it is clear now that the integral $\int
1/(2\kappa +S)$ diverges. The constant $C$ in (\ref{solution}) must then
vanish. In the presence of a finite cutoff $L_{u}$, there will then be a
very long intermediate in time asymptotic where for scales $\ll L_{u}$ the
behavior without cutoff is observed. However, after a very large time ${\cal %
T}_{L_{u}}\sim (L_{u}/\kappa )^{2}$ the dynamics changes\thinspace : the
inverse cascade stops, $\langle \theta ^{2}\rangle $ saturates to a finite
value, the system starts dissipating a finite amount of energy in the limit
of large ${\rm Pe}$ and the correlation function $\Omega (x)$ tends to the
solution (\ref{solution}) with $C=0$. The finite contribution $-\int \chi
/(2\kappa +S)$ needed to ensure the zero integral condition comes from a
strip of negative values that becomes more and more extended with time and
whose amplitude tends to vanish. More details on the dynamics of the pair
correlation function of the scalar gradients at infinite times in the
presence of an infrared cutoff $L_{u}$ may be found in Appendix \ref
{s:Longtime}.

\section{Dynamics in particle (Lagrangian) formalism}

\label{s:lagrangian}

In this Section we shall discuss the dynamical formulation of the equations
of motion. The goal is the same as in \cite{95CFKLa,94CGK}\thinspace :
reduce the calculation of simultaneous scalar statistics to averaging of
functionals of the random-in-time strain-vorticity matrix. This reduction is
crucially based on the fact that the velocity field is smooth ($\gamma =0$)
and can be performed for any smooth flow, independently of its
compressibility and the dimension of space. In the specific 1d case, no
matrices are obviously involved and one is left with averaging of a single
scalar field. Compressibility makes however the derivation slightly more
involved and some care must be taken in the ordering of the advective term.
This emerges, in particular, in the non-vanishing average of $\sigma $ in
the weight (\ref{q6b}) for the Lagrangian trajectories (\ref{05}). The
dynamical formulation is derived here using particle formalism. The
equivalent derivation using Martin-Siggia-Rose field formalism is reserved
for the Appendix.

Equation (\ref{original}) for the passive scalar can be presented in the
form 
\begin{equation}
\theta (T;x)=\int\limits_{0}^{T}\mbox{T}\exp \left[ \int\limits_{t}^{T}\hat{%
{\cal P}}(t^{\prime };x)dt^{\prime }\right] \phi
(t;x)dt=\int\limits_{0}^{T}dt\int dy\Psi (t,T;x,y)\phi (t;y),  \label{2a}
\end{equation}
where it was supposed that no pumping was supplied at negative times. In (%
\ref{2a}), the operator ${\cal P}(t;x)\equiv - u(t;x)\partial _{x}+\kappa
\partial _{x}^{2}$, time-ordered exponential is denoted by $\mbox{T}\exp $
and the function $\Psi $ can be expressed by using Lagrangian trajectories
as 
\begin{equation}
\Psi (t,T;y,x)=\int\limits_{\rho (t)=y}^{\rho (T)=x}{\cal D}\rho \exp \left[
-\int_{t}^{T}\left( \dot{\rho}-u\right) ^{2}/4\kappa \right] .
\label{interm}
\end{equation}
This formula expresses the simple fact that Lagrangian trajectories are
fixed by the velocity $u$ and smoothed by the molecular diffusivity $\kappa $%
. One can express (\ref{interm}) in the more convenient Hamiltonian form 
\begin{eqnarray}
&&\Psi (t,T;x,y)\equiv \int\limits_{\rho _{N}=x}^{\rho
_{0}=y}\prod\limits_{n=0}^{N-1}dp_{n}\prod\limits_{n=1}^{N- 1}d\rho _{n}\exp
\left[ \Delta \sum\limits_{n=1}^{N}\left[ \frac{1}{\Delta }p_{n- 1}(\rho
_{n}-\rho _{n-1})-p_{n-1}u(t_{n};\rho _{n})+\kappa p_{n- 1}^{2}\right]
\right] ,  \label{2b} \\
&&\rho _{n}=\rho (t_{n}=t+n\Delta ),\quad p_{n}=p(t+n\Delta ),\quad \Delta
\equiv \frac{T-t}{N-1}\to 0,  \label{2d}
\end{eqnarray}
where the momenta integrations ($dp_{k}$) run along the imaginary axis and
regularizations have been specified. Using the property that both velocity
and pumping are Gaussian and have correlation functions (\ref{corr}) and (%
\ref{forza}), we can easily perform the averages in the $2n$-th simultaneous
product of the scalar field $\theta (T;x)$. We thus obtain 
\begin{eqnarray}
F(T;x_{1},\cdots ,x_{2n}) &\equiv &\langle \theta (T,x_{1})\cdots \theta
(T,x_{2n})\rangle  \nonumber \\
&=&\int\limits_{0}^{T}dt\int \prod\limits_{i=1}^{2n}dy_{i}{\cal R}%
(T-t;x_{i},y_{i})\left[ F(t;y_{1},\cdots ,y_{2n-2})\chi (y_{2n}- y_{2n-1})+%
\mbox{permutations}\right] ,  \label{Ff}
\end{eqnarray}
where the eddy-diffusivity resolvent is defined as 
\begin{eqnarray}
{\cal R}(T;x_{i},y_{i}) &\equiv &\prod\limits_{i=1}^{2n}\Biggl\langle\Psi
(T,0;x_{i},y_{i})\Biggr\rangle  \nonumber \\
&=&\int\limits_{\rho _{i}(0)=x_{i}}^{\rho _{i}(T)=y_{i}}{\cal D}\rho _{i}(t)%
{\cal D}p_{i}(t)\exp \left( \int\limits_{0}^{T}dt\left[ p_{i}\dot{\rho}_{i}-%
\frac{1}{2}p_{i}(\rho _{i}-\rho _{j})^{2}p_{j}+\kappa p_{i}^{2}\right]
\right) .  \label{q1}
\end{eqnarray}
Here, one inverses the direction of time in comparison with (\ref{2b}) ( $%
t\rightarrow T-t$) and thus the convective term ($p^{2}\rho ^{2}$) is
regularized in a way such that its $\rho $-dependent part is retarded with
respect to the $p$-dependent one. The Hubbard-Stratonovich transformation of
the diffusive term gives 
\begin{eqnarray}
&&{\cal R}(T;x_{i},y_{i})=\int\limits_{\rho _{i}(0)=x_{i}}^{\rho
_{i}(T)=y_{i}}{\cal D}\rho _{i}(t){\cal D}p_{i}(t){\cal D}\xi _{i}(t)\exp
\left[ -{\cal S}-{\cal S}_{\xi }\right] ,  \label{q2a} \\
&&{\cal S}=\int\limits_{0}^{T}dt\left[ - p_{i}\dot{\rho}_{i}+\frac{1}{2}%
p_{i}(\rho _{i}-\rho _{j})^{2}p_{j}-p_{i}\xi _{i}\right] ,\quad {\cal S}%
_{\xi }=\frac{1}{4\kappa }\int\limits_{0}^{T}\xi _{i}^{2}dt.  \label{q2b}
\end{eqnarray}

In the integration in (\ref{Ff}) of the resolvent, the correlation function $%
F$ and the pumping correlation $\chi $ appear. Both depend on the difference
of the coordinates only. This means that one can simply integrate with
respect to collective variables, say $\rho =\sum \rho _{i}$ and the
collective momentum $P\equiv \sum_{i}^{2n}p_{i}/(2n)$. To get rid of the
superfluous degrees of freedom, let us come from the old set of variables $%
\{\rho _{1},\cdots ,\rho _{2n};p_{1},\cdots ,p_{2n}\}$ to the new one $%
\{\rho ,\tilde{\rho}_{2},\cdots ,\tilde{\rho}_{2n};P,\tilde{p}_{2},\cdots ,%
\tilde{p}_{2n}\}$. The new momenta $\tilde{p}_{i}=p_{i}-P$ are considered in
the system commoving with the ''center of mass'' and the positions $\tilde{%
\rho}_{i}=\rho _{i}-\rho _{1}$ are with respect to one, e.g. $\rho _{1}$,
taken as reference. The action $S$ can then be decomposed as ${\cal S}={\cal %
S}_{col}+\tilde{S}$, where 
\begin{eqnarray}
&&{\cal S}_{col}\equiv \int\limits_{0}^{T}dt\left[ \!- \!P^{2}\left(
\sum\limits_{i>1}\tilde{\rho}_{i}\right) ^{2}\!+\!2nP^{2}\left(
\sum\limits_{i>1}\tilde{\rho}_{i}^{2}\right) \!+\!2nP\left( \sum\limits_{i>1}%
\tilde{p}_{i}\tilde{\rho}_{i}^{2}\right) -2P\left( \sum\limits_{i>1}\tilde{p}%
_{i}\tilde{\rho}_{i}\right) \left( \sum\limits_{j>1}\tilde{\rho}_{j}\right)
\!-\!P\dot{\rho}\!-\!P\sum_{i=1}\xi _{i}\right] ,  \label{q4b} \\
&&\tilde{{\cal S}}\equiv \int\limits_{0}^{T}dt\left[ -\left( \sum_{i>1}%
\tilde{p}_{i}\tilde{\rho}_{i}\right) ^{2}- \sum_{i>1}\tilde{p}_{i}\dot{%
\tilde{\rho}_{i}}-\sum_{i>1}\tilde{p}_{i}(\xi _{i}-\xi _{1})\right] .
\label{q4c}
\end{eqnarray}

Let us now consider the integral $\int \prod\limits_{i=1}^{2n}dy_{i}{\cal R}%
(T;x_{i},y_{i})f(y_{i})$ where, as in (\ref{Ff}), $f$ is a function of
differences only $f(y_{i})=f(y_{i}-y)$. Collective degrees of freedom $\rho $
and $P$ are easily integrated. The principal point for this integration is
the absence of a ``potential'' $\rho $ dependence in the action ${\cal S}$.
The integral is then reduced to $\int \prod\limits_{i>1}^{2n}d\tilde{y}_{i}%
{\cal R}(T;\tilde{x}_{i},\tilde{y}_{i})f(\tilde{y}_{i}+y_{1})$, where the
effective resolvent 
\begin{equation}
\tilde{{\cal R}}(T;\tilde{x}_{i},\tilde{y}_{i})=\int\limits_{\tilde{\rho}%
_{i}(0)=\tilde{x}_{i}}^{\tilde{\rho}_{i}(T)=\tilde{y}_{i}}{\cal D}\tilde{\rho%
}(t){\cal D}\xi _{i}(t){\cal D}\tilde{\sigma}(t)\exp \left[ - {\cal S}%
_{\sigma }-{\cal S}_{\xi }\right] \prod_{m,i}\delta \left( \frac{\rho
_{i}^{(m)}-\rho _{i}^{(m-1)}}{\Delta }-\tilde{\sigma}^{(m)}\rho
_{i}^{(m-1)}-\xi _{i}^{(m)}+\xi _{1}^{(m)}\right) ,  \label{q5b}
\end{equation}
and the measure of averaging ${\cal S}_{\sigma }=\sum_{m}\left[ \tilde{\sigma%
}^{(m)}\right] ^{2}/4$. In order to derive (\ref{q5b}), we have decomposed
the quadratic over $\tilde{p}\tilde{\rho}$ term by means of
Hubbard-Stratonovich trick, introducing an additional collective integration
over $\tilde{\sigma}$. The integration with respect to momentum $\tilde{p}%
_{i}$ is already performed in the last line of (\ref{q5b}) (the effective
action appears to be linear in $\tilde{p}$ in the result of the
Hubbard-Stratonovich transformation). The continuous versions\footnote{%
By continuous version we understand, particularly, a symmetrical smearing of
the temporal $\delta $-function in the $\sigma $-field correlation for a
small but finite (which is still much larger than the temporal slice $\Delta 
$) width.} of the equation under the $\delta $-function sign in (\ref{q5b})
and the measure of averaging over $\tilde{\sigma}$ are 
\begin{eqnarray}
&&\dot{\tilde{\rho}_{i}}=\sigma \tilde{\rho}_{i}+\xi _{i}-\xi _{1},\quad
\sigma \equiv \tilde{\sigma}-1,  \label{q6a} \\
&&{\cal S}_{\sigma }=\frac{1}{4}\int\limits_{0}^{T}dt\left[ \sigma +1\right]
^{2}.  \label{q6b}
\end{eqnarray}
To see the relation between (\ref{q5b}) and (\ref{q6a},\ref{q6b}) one can
check, particularly, that both discrete and continuous versions give $%
\langle \rho _{i}(t)\rangle =\rho _{i}(0)$. The formal solution of (\ref{q6a}%
) is 
\begin{equation}
\tilde{\rho}_{i}(t)=W(t)(x_{i}- x_{1})+W(t)\int\limits_{0}^{t}dt^{\prime
}W^{-1}(t^{\prime })\left( \xi _{i}-\xi _{1}\right) ,\quad W(t)\equiv \exp
\left[ \int\limits_{0}^{t}dt^{\prime }\sigma (t^{\prime })\right] .
\label{05}
\end{equation}

It is finally easy to recalculate the $2n$-particle correlation function of
the scalar from the $2n$-particle resolvent 
\begin{eqnarray}
F(T;x_{1},\cdots ,x_{2n}) &=&\Biggl\langle \Xi [T;\{\sigma
(t)\};x_{1}-x_{2}]\cdots \Xi [T;\{\sigma (t)\};x_{2n-1}-x_{2n}]+%
\mbox{permutations}\Biggr\rangle_{\sigma },  \label{02} \\
&&\Xi [T;\{\sigma (t)\};x_{i}-x_{j}]\equiv \int\limits_{0}^{T}dt%
\Biggl\langle \chi \left[ W(t)\frac{x_{i}-x_{j}}{L}+W(t)\int%
\limits_{0}^{t}dt^{\prime }W^{-1}(t^{\prime })\frac{\xi _{i}-\xi _{j}}{L}%
\right] \Biggr\rangle_{\xi _{i,j}},  \label{04}
\end{eqnarray}
where averages over $\sigma (t)$ and $\xi _{i}(t)$ are fixed by the measures 
$\exp [-{\cal S}_{\sigma }]$ and $\exp [-{\cal S}_{\xi }]$ defined in (\ref
{q6b}) and (\ref{q2b}), respectively.

\section{Correlation functions of scalar gradients in the convective interval
}

\label{s:gradients}

Using the results of the previous Section, the dynamical expression of
correlation functions of the scalar gradient $\omega (t,x)=\partial
_{x}\theta (t,x)$ can be simply found differentiating (\ref{02}) with
respect to all spatial arguments. Here, we shall be interested in the
behavior of these simultaneous correlation functions for distances such that
molecular diffusivity can be neglected. We first derive a general formula
valid for arbitrary form of forcing correlation and then treat more
specifically the case of an exponential pumping. The resulting expression
shows that the ratio between the irreducible and the reducible contributions
to the $2n$-th correlation function grows as $\left( L/x\right) ^{n-1}$.
This evidences both the non-Gaussianity of the field and the fact that it
increases going downscales in the convective interval. The same phenomenon
takes place for generic forms of the pumping. The discussion of the range of
scales where these considerations hold, is postponed to the end of the
Section.

It follows from (\ref{02}) that scalar gradients correlation functions are
given by 
\begin{eqnarray}
&&\langle \omega (T,x_{1})\cdots \omega (T,x_{2n})\rangle  \nonumber \\
&=&{\frac{(-1)^{n}}{L^{2n}}}\Biggl\langle \left[
\int\limits_{0}^{T}dtW^{2}(t)\chi ^{\prime \prime }\left[ W(t)\frac{%
x_{1}-x_{2}}{L}\right] \right] \cdots \left[
\int\limits_{0}^{T}dtW^{2}(t)\chi ^{\prime \prime }\left[ W(t)\frac{%
x_{2n}-x_{2n-1}}{L}\right] \right] \Biggr\rangle+\mbox{permutations},
\label{co1}
\end{eqnarray}
where molecular diffusion effects have been neglected. The only averaging
left in (\ref{co1}) is then with respect to the $\sigma $ statistics defined
by (\ref{q6b}). A possible way of performing this average is to introduce
the auxiliary object 
\begin{equation}
A(s_{1,2},\cdots ,s_{2n,2n-1})=\Biggl\langle \exp \left[
\int\limits_{0}^{T}W^{2}(t)\left( s_{1,2}\chi ^{\prime \prime }\left[ W(t)%
\frac{x_{1}-x_{2}}{L}\right] +\cdots +s_{2n,2n-1}\chi ^{\prime \prime
}\left[ W(t)\frac{x_{2n}-x_{2n-1}}{L}\right] \right) dt\right] \Biggr\rangle.
\label{AAA}
\end{equation}
Differentiating (\ref{AAA}) over $s$ variables, (\ref{co1}) is clearly
reproduced up to the sign and the $L$ dependent prefactor. Inserting the
weight (\ref{q6b}) into (\ref{AAA}), one can easily recognize the
path-integral structure associated with the time evolution of the ``quantum
mechanical'' Schr\"{o}dinger equation, having Hamiltonian $\hat{H}=-\partial
_{\eta }^{2}-U(\exp \eta )\exp (2\eta )$. The ``potential'' $U$ appearing in
the Hamiltonian is 
\begin{equation}
U(y)\equiv s_{1,2}\chi ^{\prime \prime }\left[ y\frac{x_{1}- x_{2}}{L}%
\right] +\cdots +s_{2n,2n-1}\chi ^{\prime \prime }\left[ y\frac{x_{2n}-
x_{2n-1}}{L}\right] ,  \label{co5}
\end{equation}
and the ``space'' variable $\eta $ is defined as $\eta
=\int\limits_{0}^{t}dt^{\prime }\sigma (t^{\prime })$. Using standard
notation for quantum mechanics matrix elements, it is easy to check that (%
\ref{AAA}) can be presented as 
\begin{equation}
A=\exp [-T/4]\Biggl\langle\delta (\eta )\Biggl|e^{- T\hat{H}}\Biggr| %
e^{-\eta /2}\Biggr\rangle=\left[ e^{-T/4}\Pi (T;\eta )\right] _{\eta =0},
\end{equation}
where the ''wave-function'' $\Pi (T;\eta )$ satisfies $(\partial _{T}-\hat{H}%
)\Pi =0$, and the initial condition is $\Pi (0;\eta )=e^{-\eta /2}$. We can
now remark that the potential part of the Hamiltonian vanishes at $\eta \to
-\infty $, while the initial condition does not. The resulting asymptotic
behavior at large times $T$ will then be 
\begin{equation}
\Pi (T;\eta =\ln [y])\to \Biggl|_{T\to \infty }e^{T/4}\frac{\Pi _{0}}{\sqrt{y%
}},\qquad \left[ \partial _{y}^{2}+U(y)\right] \Pi _{0}(y)=0.  \label{co6}
\end{equation}
The new variable $y=\exp \eta $ has been introduced. The boundary conditions
for the spatial part $\Pi _{0}$ are easily derived from those for $\Pi $%
\thinspace : it should tend to unity for $y\to 0$ and the ratio $\Pi _{0}/%
\sqrt{y}$ should vanish for $y$ tending to infinity. It follows from (\ref
{AAA}) that the auxiliary object $A$ is simply the function $\Pi _{0}$
calculated at $y=1$.

Derivatives of $A$ at $s=0$ are needed for the calculation of gradients
correlations (\ref{co1}). It is then convenient to present the solution of (%
\ref{co6}) in the form of an expansion with respect to the ``potential'' $U$%
\thinspace : 
\begin{equation}
A(s_{1,2},\cdots ,s_{2n,2n-1})=\sum\limits_{k=0}^{\infty
}\int\limits_{0}^{1}dy_{1}\int\limits_{y_{1}}^{\infty
}dy_{2}U[y_{2}]\int\limits_{0}^{y_{2}}dy_{3}\int\limits_{y_{3}}^{\infty
}dy_{4}U[y_{4}]\cdots
\int\limits_{0}^{y_{2k-2}}dy_{2k-1}\int\limits_{y_{2k-1}}^{\infty
}dy_{2k}U[y_{2k}].  \label{co8}
\end{equation}
Only the $n$-th order term of the expansion in (\ref{co8}) actually
contributes to the $2n$-th order scalar gradients correlation function. Its
final expression reads 
\begin{eqnarray}
&&\langle \omega _{1}\cdots \omega _{2n}\rangle ={\frac{(- 1)^{n}}{L^{2n}}}%
\sum \int\limits_{0}^{1}dy_{1}\int\limits_{y_{1}}^{\infty }dy_{2}\chi
^{\prime \prime }\left[ y_{2}\frac{x_{k_{1}}-x_{k_{2}}}{L}\right]
\int\limits_{0}^{y_{2}}dy_{3}\int\limits_{y_{3}}^{\infty }dy_{4}\chi
^{\prime \prime }\left[ y_{4}\frac{x_{k_{3}}-x_{k_{4}}}{L}\right] \cdots 
\nonumber \\
&&\times \int\limits_{0}^{y_{2n-2}}dy_{2n-1}\int\limits_{y_{2n- 1}}^{\infty
}dy_{2n}\chi ^{\prime \prime }\left[ y_{2n}\frac{x_{k_{2n}}- x_{k_{2n-1}}}{L}%
\right] ,  \label{co9}
\end{eqnarray}
where summation is performed over all the splittings of the set $%
\{k_{1},\cdots ,k_{2n}\}$ into $n$ ordered pairs.

Formula (\ref{co9}) holds for any pumping $\chi$. Let us now specifically
consider the case of the exponential pumping $\chi _{e}(z)=\exp [-|z|]$,
where the integrals appearing in (\ref{co9}) can be easily performed. For
the second-order correlation, we obtain that the dominant contribution for
small distances $x_{12}/L$ is $-1/(x_{12}L)$, in agreement with the solution
(\ref{solution}). For the fourth-order correlation function, we find 
\begin{equation}
\langle \omega(x_1,t)\omega(x_2,t)\omega(x_3,t)\omega(x_4,t)\rangle
_{e}=F_{e}(x_{12},x_{34})+F_{e}(x_{13},x_{24})+F_{e}(x_{14},x_{23}), \qquad
F_{e}(x,y)\simeq \frac{2}{xy(x+y)L},  \label{co10c}
\end{equation}
where $x_{ij}\equiv |x_{i}-x_{j}|$, the subscript is intended to remind that
this specific expression holds for the exponential pumping and only dominant
terms in $x_{ij}/L$ have been retained. Distinguishing between reducible
(Gaussian) and irreducible contributions into (\ref{co10c}), one observes
that the irreducible part is $L/x\gg 1$ times larger. More generally, if $x$
denotes the typical distance among the various particles, i.e. $x_{ij}\sim x$%
, then $\langle \omega _{1}\cdots \omega _{2n}\rangle \sim 1/\left(x^{2n-
1}L\right)$. Both the correlation functions and the degree of
non-Gaussianity of the scalar gradient (ratio of the $2n$-th moment to its
reducible part) are then growing with $x_{ij}$ going downscales.

Let us finally come to the range of validity of the previous convective
arguments. The exponential form of the pumping is a special one, being not
regular at the origin. The first term of its expansion at small distances is
linear and not quadratic. This affects the dependence on $x$ of the
second-order correlation. For a regular pumping, the dominant term would
indeed be constant, as follows from (\ref{small}). On the contrary, one can
check using (\ref{co9}) that the dependence of correlations of order $\geq 4$
on $x$ and $L$ remains the same as for the exponential pumping. The
regularity at the origin of the pumping also enters the range of scales
where neglecting molecular diffusivity effects is allowed. Performing the
small distance expansion, as in (\ref{small}) for a regular pumping, a
logarithmic correction $\propto \log {\rm Pe}$ appears. The ultraviolet
criterium of applicability of the previous convective considerations for the
exponential pumping is then $x\gg \sqrt{\kappa }\ln [\mbox{Pe}]$. For a
pumping regular at the origin, the criterium is the same as in (\ref{small}%
), i.e. $x\gg L/\sqrt{{\rm Pe}}$.

\section{Scaling and PDF of scalar differences}

\label{s:differences}

Scalar structure functions $S_{2n}(x)=\langle (\theta (T,x)- \theta
(T,0))^{2n}\rangle $ can be easily expressed in terms of scalar correlation
functions $F$ by taking the appropriate combinations of them. A dynamical
expression for $S_{2n}$ can thus be derived directly from (\ref{04}),
obtained in Section~\ref{s:lagrangian} for the $F$'s. This is however not a
very practical procedure. Each of the contributions, appearing in the sum
giving the structure functions, contains indeed the constant mode. As it was
discussed in Section~\ref{ss:model} and as it also appears from the
dynamical expression (\ref{02}), this mode grows linearly with the
observation time $T$. It is just in the whole sum that these divergent
contributions are canceled, thus leaving the time-independent final result
for structure functions. It is then more convenient to restore structure
functions directly from scalar gradients correlations as $%
S_{2n}(x)=\int\limits_{0}^{x}dx_{1}\cdots \int\limits_{0}^{x}dx_{2n}\langle
\omega _{1}\cdots \omega _{2n}\rangle $. An important question is whether,
to calculate $S_{2n}$ in the convective interval, we can avoid to take
dissipation explicitly into account or not. This means essentially asking
whether it is enough to know just the convective expressions for gradients
correlations or their whole behavior is needed. This point can be tested by
simply taking the convective expressions for $\langle \omega _{1}\cdots
\omega _{2n}\rangle $ found in the previous Section and inserting them into
the integral expression for $S_{2n}$. One can then check that all the
integrals for any structure function are convergent on the ultraviolet and
dominated by the infrared side of the convective range.

The expression for structure functions in the convective interval is then 
\begin{eqnarray}
&&S_{2n}(x)=  \nonumber \\
&&(2n- 1)!!n!2^{n}\int\limits_{0}^{x/L}dy_{1}\int\limits_{y_{1}}^{\infty
}dy_{2}\frac{\chi [0]-\chi [y_{2}]}{y_{2}^{2}}\int\limits_{0}^{y_{2}}dy_{3}%
\int\limits_{y_{3}}^{\infty }dy_{4}\frac{\chi [0]-\chi [y_{4}]}{y_{4}^{2}}%
\cdots \int\limits_{0}^{y_{2n-2}}dy_{2n-1}\int\limits_{y_{2n- 1}}^{\infty
}dy_{2n}\frac{\chi [0]-\chi [y_{2n}]}{y_{2n}^{2}},  \label{sd3}
\end{eqnarray}
where we have already performed the $2n$ integrals over the $dx_{i}$'s. The
whole set of equations (\ref{sd3}) can be recast into the more compact
equation 
\begin{equation}
\left(x^{2}\partial_{x}^2 -\lambda ^{2}\left[ \chi (0)-\chi ({\frac{x}{L}}%
)\right] \right) {\cal Z}({\frac{x}{L}},\lambda )=0,  \label{sd5}
\end{equation}
for the generating function ${\cal Z}(x,\lambda )=\left\langle \exp \left(
-i\lambda \delta \theta _{x}\right) \right\rangle$ of scalar differences $%
\delta \theta _{x}$ taken at the scale $x$. From this very definition, it
follows that the function $Z$ must tend to unity for vanishing $x$ and, for
the convergence of the integrals in (\ref{sd3}), it should grow slower than
linearly at infinity.

It is worth to remind that (\ref{sd5}) was found as the result of an
accurate dynamical procedure\thinspace : we first calculated correlation
functions of the scalar gradient for all points being separated\thinspace ;
the resulting gradients correlation functions were then integrated to obtain
structure functions and, finally, the generating function for scalar
differences was reconstructed from its moments. We thus avoided to
explicitly handle diffusion, paying for this the price of taking many
particles into consideration. The closed differential equation (\ref{sd5})
for scalar differences generating function emerges as the result of this
procedure. On the other hand, one could generally (also for the non-smooth
case, i.e. $\gamma \neq 0$ in (\ref{corr})) derive the following unclosed
Fokker-Planck equation for the generating function 
\begin{equation}
\left( x^{2-\gamma }\partial _{x}^{2}-\lambda ^{2}\left[ \chi (0)-\chi
(x/L)\right] \right) {\cal Z}(x/L,\lambda )=\kappa \Biggl\langle\left[
\partial _{1}^{2}\theta _{1}-\partial _{2}^{2}\theta _{2}\right] \exp \left[
\lambda [\theta _{1}-\theta _{2}]\right] \Biggr\rangle.  \label{sd5u}
\end{equation}
This equation is simply obtained averaging the equation of motion (\ref
{original}) for the scalar at two reference points. The smooth ($\gamma =0$)
limit of (\ref{sd5u}) differs from (\ref{sd5}) by the r.h.s. dissipative
term. There is a general expectation that this term may remain finite even
in the limit $\kappa \to 0$, thus providing a nonvanishing anomaly in the
terminology of Polyakov \cite{95Pol}. Equation (\ref{sd5}), derived
microscopically without any conjecture, shows then the absence of anomaly
for the one dimensional smooth flow\footnote{%
We acknowledge A. Polyakov for attracting out attention to the matter
discussed in the paragraph.}. Physically, the absence of anomaly is
associated with the vanishing direct flux of $\theta ^{2}$ in the limit of
infinite P\'{e}clet numbers. This point, already appeared in Section~\ref
{ss:model}, will emerge even more clearly in the full analysis of the
dissipation field in the next Section. In the two following Subsections,
solutions of (\ref{sd5u}) at $x\gg L$ and $x\ll L$ will be discussed.

\subsection{Upscale interval}

Let us first consider scales larger than the integral scale $L$. The
asymptotic solution for the generating function, which can also be found
from (\ref{sd5u}) replacing $\chi (0)- \chi (x/L)$ by $\chi (0)$, is 
\begin{equation}
{\cal Z}=\left( \frac{x}{L}\right) ^{1/2-\sqrt{1+4\chi [0]\lambda ^{2}}/2}.
\label{x>L}
\end{equation}
The inverse Fourier transform of (\ref{x>L}), calculated in the saddle-point
manner (the large parameter is $x/L$ or, equivalently, large values of $%
|\delta \theta _{x}|$), gives 
\begin{equation}
{\cal P}^{\delta \theta _{x}}(y)=\frac{1}{2\sqrt{\pi \chi [0]\ln [x/L]}}%
\left\{ 
\begin{array}{cc}
\exp \left[ -y^{2}/(4\chi [0]\ln [x/L])\right] ,\quad & |y|\ll \ln [x/L], \\ 
\exp \left[ -|y|/(2\sqrt{\chi [0]})\right] ,\quad & |y|\gg \ln [x/L].
\end{array}
\right.  \label{pdfx>L}
\end{equation}
Thus, the arguments for Gaussianity presented in \cite{95CFKLa} for the
incompressible case can indeed be reversed and applied here for the upper
interval. Structure functions of orders much less than $\ln [x/L]$ scale
indeed logarithmically, the core of scalar differences PDF is Gaussian and
the PDF's tail is exponential.

\subsection{Downscale interval}

We shall now obtain a general formula for the PDF at $y\ll 1$, no matter how 
$y$ and $x/L$ relate to each other, provided both of them are small. Indeed,
replacing the $\chi $ dependent ``potential'' by the first term of the
expansion over $x/L$, and solving the resulting differential equation with
the same boundary conditions as before, one gets the simple exponential form
for the generating function ${\cal Z}(x/L,\lambda )=\exp [- \lambda \sqrt{%
\chi ^{\prime \prime }[0]/2}\,\,x/L]$. The inverse Fourier transform
produces the following Lorentzian expression for the PDF 
\begin{equation}
{\cal P}^{(\delta \theta _{x})}(y)=\frac{1}{\pi }\frac{L}{x}\frac{1}{%
y^{2}+|\chi ^{\prime \prime }[0]|\,x^{2}/(2L^{2})},  \label{Lor}
\end{equation}
which is thus generally valid at $|y|\ll 1$. It results from (\ref{Lor})
that the PDF is smooth in a small region around the origin $y=0$, where it
can actually be expanded in $y^{2}L^{2}/x^{2}$. This region extends
approximately up to $x/L$, where the second behavior in $1/y^{2}$ sets in.
Note that the concavity of the PDF in this second region is upwards and
remain upwards up to the small values $x/L$. It is just for very small
values $|y|\ll x/L $ that the concavity is reversed downwards. An
``experimental'' histogram of such a PDF would then look strongly cusped at
the origin.

The tail of the PDF matching the Lorentzian (\ref{Lor}) at $|y|\sim 1$ is
exponential. To see this, and generally to obtain an explicit analytic
solution for the PDF in the whole domain of $x$ and $\delta \theta _{x}$,
let us consider the following specific form of pumping 
\begin{equation}
\quad \chi _{*}({\frac{x}{L}})=\frac{1}{1+\left( x/L\right) ^{2}}.
\label{sd7}
\end{equation}
Making in (\ref{sd5}) the change of variable $\cot [\varphi ]=x/L$, the
solution of the resulting equation can be expressed in terms of associated
Legendre functions as 
\begin{equation}
{\cal Z}_{*}(\cot [\varphi ];\lambda )=\frac{2^{\nu -1/2}(\nu +1)\Gamma
^{2}\left[ (\nu +1)/2\right] P_{1/2}^{-\nu -1/2}\left( \cos [\varphi
]\right) }{\sqrt{\pi \sin [\varphi ]}},  \label{sd80}
\end{equation}
where the upper index $\nu =-1/2+1/2\,\sqrt{1+4\lambda ^{2}}$. The choice of
the sign for the square root is such as to ensure that the generating
function $Z(x,\lambda )$ grows at infinity slower than linearly. The
notation ${\cal Z}_{*}$ in (\ref{sd80}) is intended to stress that this
explicit solution was obtained for the pumping (\ref{sd7}). Note, that (\ref
{sd80}) is particularly applicable for the upper interval discussed in the
previous Subsection. Considering (\ref{sd80}) at $\varphi \ll 1$ and $%
\lambda \gg 1$, one indeed recovers (\ref{x>L}). Using the integral
representation (8.714) in \cite{GR} and the doubling formula for the gamma
function, (\ref{sd80}) can be presented in the integral form 
\begin{equation}
{\cal Z}_{*}(\cot [\varphi ];\lambda )=\frac{2\Gamma [(\nu +3)/2]}{\sqrt{\pi 
}\sin [\varphi ]\Gamma [(\nu +2)/2]}\int\limits_{0}^{\varphi }\cos [t]\left( 
\frac{\cos [t]-\cos [\varphi ]}{\sin [\varphi ]}\right) ^{\nu }\,dt.
\label{sd8}
\end{equation}
The integral representation (\ref{sd8}) is useful, since it clearly shows
the analytic structure of ${\cal Z}_{*}$ with respect to $\lambda $. One
observes, particularly, that ${\cal Z}_{*}$ is analytic in the whole upper
semi-plane, except on the line of imaginary $\lambda $ from $i/2$ to $%
i\infty $. The path on the real axis in the inverse Fourier transform can be
then deformed into the following one surrounding the cut\thinspace : the
left branch goes from $i\infty -0^{+}$to $i/2-0^{+}$, bypass of $\lambda
=i/2 $ from below and the right branch goes from $i/2+0^{+}$ to $i\infty
+0^{+}$. The final result is 
\begin{equation}
{\cal P}_{*}^{(\delta \theta _{x})}(y)=\frac{1}{4\pi }\int\limits_{0}^{%
\infty }\frac{qdq}{\sqrt{1+q^{2}}}\exp \left[ - \frac{|y|}{2}\sqrt{(1+q^{2})}%
\right] {\cal G}(\cot [\varphi ];q),  \label{sd10}
\end{equation}
where the function $-i{\cal G}$ is the difference between ${\cal Z}_{*}$ on
the right and the left of the cut, i.e. $\nu \to -1/2\pm iq/2$,
respectively. The general expression for ${\cal G}$ can be derived from the
integral representation (\ref{sd8}) as 
\begin{equation}
{\cal G}(\cot [\varphi ];q)=\frac{4}{\sqrt{\pi \sin [\varphi ]}}%
\int\limits_{0}^{\varphi }\frac{\cos [t]dt}{\sqrt{\cos [t]-\cos [\varphi ]}}%
\mbox{Re}\left[ i\frac{\Gamma \left( \frac{5+iq}{4}\right) }{\Gamma \left( 
\frac{3+iq}{4}\right) }\left[ \frac{\sin [\varphi ]}{\cos [t]\!- \!\cos
[\varphi ]}\right] ^{-iq/2}\right] .  \label{sd10a}
\end{equation}

The calculation of the PDF of scalar differences in the convective interval
is now reduced to the evaluation of the asymptotic behavior of ${\cal G}$ in
(\ref{sd10a}) and then to perform the integral (\ref{sd10}). In the
convective interval, $x/L=\cot [\varphi ]\ll 1$, i.e. the angle $\varphi $
is very close to $\pi /2$. The dominant expression of ${\cal G}$ in this
region can be obtained by simply expanding directly in (\ref{sd10a}).
Inserting this expansion into (\ref{sd10}), we obtain 
\begin{equation}
{\cal P}_{*}^{(\delta \theta _{x})}(y)=\frac{\pi }{8}\frac{x}{L}%
\int\limits_{0}^{\infty }\frac{q\sqrt{1+q^{2}}\sinh [\pi q/2]\exp \left[ -%
\frac{|y|}{2}\sqrt{1+q^{2}}\right] dq}{|\Gamma \left[ (3+iq)/4\right]
|^{4}\cosh ^{2}[\pi q/2]}\to \left\{ 
\begin{array}{cc}
\frac{1}{\pi }\frac{x}{L}\frac{1}{y^{2}},\quad & 1\gg |y|\gg x/L, \\ 
\sim \frac{x}{L}\exp [-|y|/2],\quad & |y|\gg 1,
\end{array}
\right. ,  \label{sd11a}
\end{equation}
where, at $y\gg 1$, the prefactor algebraic in $y$ has not been considered.
Varying the pumping function $\chi (x/L)$, one can change the number behind $%
|y|$ in the exponential, but the exponential behavior itself will never
change.

One may check that the PDF's asymptotic for the smallest values $|y|\ll x/L$
derived from (\ref{sd80},\ref{sd8}) is consistent with the general formula (%
\ref{Lor}). Indeed, the respective large $q$ asymptotic of ${\cal G}$ is 
\begin{equation}
{\cal G}(\cot [\varphi ];q)\sim 2\sqrt{\frac{q}{\pi \sin [\varphi ]}}%
\int\limits_{0}^{\varphi }\frac{\cos [t]}{\sqrt{\cos [t]-\cos [\varphi ]}}%
\sin \left( \frac{q}{2}\ln \left[ \frac{\sin [\varphi ]}{\cos [t]-\cos
[\varphi ]}\right] -\frac{\pi }{4}\right) .  \label{sd12}
\end{equation}
Substituting (\ref{sd12}) into (\ref{sd10}) and performing first the
integral in $q$ and then the one in $t$, we obtain at $|y|\ll x/L\ll 1$%
\begin{equation}
{\cal P}_{*}^{(\delta \theta _{x})}(y)\sim \frac{1}{\sqrt{2}\pi }%
\int\limits_{0}^{\varphi }\frac{\cos [t]dt}{\sqrt{\cos [t]-\cos [\varphi ]}}%
\ln ^{-3/2}\left[ \frac{1}{\cos [t]-\cos [\varphi ]}\right] \left( 1-\frac{15%
}{8}\frac{y^{2}}{\ln ^{2}\left[ \frac{1}{\cos [t]-\cos [\varphi ]}\right] }%
\right) \to \frac{1}{\pi }\frac{L}{x}\left( 1- \frac{L^{2}y^{2}}{x^{2}}%
\right)  \label{sd13a}
\end{equation}
Expressions (\ref{sd11a}) and (\ref{sd13a}) are clearly in agreement with (%
\ref{Lor}), valid for an arbitrary form of pumping.

Let us now derive the scaling behavior of scalar structure functions $%
\langle |\theta _{1}-\theta _{2}|^{a}\rangle $ at $x\ll L$. The scaling for
orders $a>1$ is dominated by the behavior of the PDF at values of order
unity. On the contrary, for $a<1$, the region in $1/y^{2}$ dominates. The
resulting scaling behavior of structure functions is 
\begin{equation}
\langle |\theta _{1}-\theta _{2}|^{a}\rangle \sim \cases{{x\over L} ,&for $a
\ge 1$;\cr \cr \left({x\over L}\right)^a ,&for $-1 < a\le 1$.\cr}
\label{Burgers}
\end{equation}
For $a<-1$ the moments do not exist at all, since the PDF is finite at the
origin. Remark that (\ref{Burgers}) is exactly the same scaling as for
velocity structure functions in Burgers equation. In the present model the
collapse of high-order exponents is associated with the uniformity in space
of stretchings and compressions.

\section{PDF of scalar dissipation and gradients}

\label{s:dissipation}

The $n$-th order moment of the dissipation field $\epsilon =\kappa \left(
\partial _{x}\theta (t;x)\right) ^{2}$ can be obtained from the dynamical
expression (\ref{04}) as 
\begin{equation}
\langle \epsilon ^{n}\rangle =(2n-1)!!\,\kappa ^{n}\,\Biggl\langle\left[
\partial _{x_{i}}\partial _{x_{j}}\Xi [T;\{\sigma (t)\};x_{i}-x_{j}]\right]
_{x_{i}=x_{j}}^{n}\Biggr \rangle \equiv (2n-1)!!\,\,\Biggl\langle %
Q[T;\{\sigma (t)\}]^{n}\Biggr\rangle_{\sigma }.  \label{06}
\end{equation}
The average with respect to the noises $\xi _{j}$ is easily performed in
Fourier space, thus obtaining 
\begin{equation}
Q\equiv \frac{1}{\mbox{Pe}^{2}}\int\limits_{0}^{T}dtW^{2}(t)G\left[ \frac{%
W^{2}(t)}{\mbox{Pe}^{2}}\int\limits_{0}^{t}dt^{\prime }W^{-2}(t^{\prime
})\right] \qquad G(x^{2})\equiv \frac{1}{4\pi }\int\limits_{-\infty
}^{\infty }dqq^{2}\chi _{q}\exp \left( -q^{2}x^{2}\right) ,  \label{07}
\end{equation}
where $W(t)$ has been defined in (\ref{04}) and $\chi _{q}$ is the Fourier
transform of the pumping correlation function $\chi (x)$. Let us remind that
the P\'{e}clet number is defined as the pumping-to-diffusive scale ratio $%
\mbox{Pe}\equiv L/\sqrt{2\kappa }$ and is supposed to be large. To obtain
the stationary value $\langle \epsilon ^{n}\rangle $, the limit $T\to \infty 
$ should be considered.

The average over $\sigma$ with the Gaussian weight (\ref{q6b}) has to be
calculated in (\ref{06}). This is very hard to do explicitly in the general
case. We can however exploit the presence of the large parameter ${\rm Pe}$
to develop an asymptotic theory that captures the dominant terms in (\ref{06}%
) with respect to ${\rm Pe}$. The important point is that, when ${\rm Pe}$
is large, there are two very different time scales in the dynamics. For the
Lagrangian trajectories relevant to (\ref{06}), particles start very close.
The additive molecular noise term in the Langevin equation for Lagrangian
trajectories is dominant at these distances and remains dominant as long as
the particles do not separate by a distance comparable to the dissipative
scale. This phase of the dynamics corresponds to the formation of the
integral in the square brackets in (\ref{07}) and takes place on times of
order unity (not scaling with ${\rm Pe}$). Once the particles have reached
the dissipative scale and enter into the convective region, random
multiplicative effects due to the velocity become dominant. Due to the
multiplicative nature of the dynamics, the time to go from the dissipative
scale to the integral scale varies as $\ln {\rm Pe}$. This phase is
associated with the growth of the $W^2$ terms in (\ref{07}). For large
P\'eclet numbers, the two processes, formation of the integral in the square
brackets and growth of $W^2$ terms in (\ref{07}), are well separated in
time. Let us then consider a time $t_0$ satisfying $1\ll t_{0}\ll \ln [%
\mbox{Pe}]$. At the dominant order in P\'eclet, $Q$ can be approximated as 
\begin{equation}
Q[T;\{\sigma (t)\}]\approx \frac{\beta }{\mbox{Pe}^{2}}\int%
\limits_{t_{0}}^{T}dt\exp \left[ 2\int\limits_{t_{0}}^{t}\sigma
_{>}(t^{\prime })dt^{\prime }\right] G\left[ \alpha \frac{\exp \left[
2\int\limits_{t_{0}}^{t}\sigma _{>}(t^{\prime })dt^{\prime }\right] }{%
\mbox{Pe}^{2}}\right] ,  \label{Q1}
\end{equation}
where $\alpha$ and $\beta$ are defined as 
\begin{equation}
\alpha \equiv \exp \left[ 2\int\limits_{0}^{t_{0}}\sigma (s)ds \right]
\int\limits_{0}^{t_{0}}dt^{\prime }\exp \left[ - 2\int\limits_{0}^{t^{\prime
}}\sigma (t^{\prime \prime })dt^{\prime \prime }\right] ,\quad \beta \equiv
\exp \left[ 2\int\limits_{0}^{t_{0}}\sigma (t^{\prime })dt^{\prime }\right]
\label{10}
\end{equation}
In order to obtain (\ref{Q1}) from the original expression (\ref{07}), we
have made the following two steps\,: $Q[t_0;\{\sigma(t)\}]$ has been
neglected and the upper bound $t$ in the integral over $dt^{\prime}$ for $%
\alpha$ has been replaced by $t_0$. Both steps are motivated by the time
scales separation at large P\'eclet numbers. More precise conditions of
validity of the approximation will be discussed later in the Section. The
moments $\langle Q^n\rangle$ can now be obtained by taking the $n$-th power
of (\ref{Q1}) and averaging. The average over $\sigma(t)$ is decomposed in a
small times part ${\cal D}\sigma_{<} \equiv
\prod\limits_{t<t_{0}}\,d\sigma(t)$ and a large times part ${\cal D}%
\sigma_{>} \equiv \prod\limits_{t>t_{0}}\,d\sigma(t)$. The corresponding
weights are simply obtained decomposing the integral over $t$ in (\ref{q6b})
as ${\cal S}_{<}=1/4\int\limits_{0}^{t_{0}}\left[ \sigma _{<}+1\right] ^{2}$
and in ${\cal S}_{>}$ the integration runs from $t_0$ to $T$. The great
advantage of (\ref{Q1}) is that, in the large times averaging, both $\alpha $
and $\beta $ are just external parameters, depending neither on time $t$ nor
on $\sigma _{>}$. Once the average over $\sigma_{>}$ is performed, we are
then left with a function of $\alpha$ and $\beta$. This is, on his turn,
averaged over $\sigma_{<}$ giving the final result $\langle Q^n\rangle$.

A compact way for averaging over large times statistics is to introduce the
Laplace transform of the PDF ${\cal P}_{s}^{>}\equiv \langle \exp
[-sQ]\rangle _{>}$. It is indeed easy to recognize that its path integral
coincides with a matrix element in the ''Quantum mechanics'' with
Hamiltonian $\hat{H}=-\partial _{\eta }^{2}+s\beta \exp \left( 2\eta \right)
G\left[ \alpha \exp \left( {2\eta }\right) \,/\mbox{Pe}^{2}\right] /\mbox{Pe}%
^{2}$. The ``space'' variable $\eta =\int_{t_{0}}^{t}\sigma (t^{\prime
})\,dt^{\prime }$. Using standard quantum mechanical notation, the
expression for ${\cal P}_{s}^{>}$ can be presented as 
\begin{equation}
{\cal P}_{s}^{>}\equiv \langle \exp [-sQ]\rangle _{>}=\exp \left[
-(T-t_{0})/4\right] \Biggl\langle\delta (\eta )\Biggl|e^{-(T- t_{0})\hat{H}%
_{0}}\Biggr|e^{-\eta /2}\Biggr\rangle =\left[ e^{-(T- t_{0})/4}\Phi
(T-t_{0};\eta )\right] _{\eta =0},  \label{P>1}
\end{equation}
where $\Phi (t;\eta )$ satisfies $(\partial _{t}-\hat{H})\Phi =0$. The
initial condition at $t=0$ for the ``wave-function'' $\Phi (t;\eta )$ is $%
\exp \left( -\eta /2\right) $. Noting that the potential part of the
Hamiltonian vanishes at $\eta \to -\infty $ and $\Phi (0;\eta )$ does not,
we obtain the asymptotic behavior of $\Phi $ at large times 
\begin{equation}
\Phi (t,\eta =\ln [y\mbox{Pe}])\to \Biggl|_{t\to \infty }e^{t/4}\frac{\Phi
_{0}(y)}{\sqrt{\mbox{Pe}\ y}},\qquad \left[ \partial _{y}^{2}- s\beta
G[\alpha y^{2}]\right] \Phi _{0}(y)=0.  \label{P>4}
\end{equation}
The new variable $y=\left( \exp \eta \right) /{\rm Pe}$ has been introduced.
The function $\Phi _{0}(y)$ should tend to unity for $y\to 0$ and $\Phi _{0}/%
\sqrt{y}$ should vanish for $y\to \infty $. It follows from (\ref{P>1}) that 
${\cal P}_{s}^{>}$ is simply the function $\Phi _{0}$ calculated at $y=1/%
{\rm Pe}$.

The general way to attack (\ref{P>4}) for an arbitrary form of pumping is to
proceed as we have already done for (\ref{co8}) in Section~\ref{s:gradients}%
. Starting with a constant unit solution, the term with $s$ in (\ref{P>4})
is treated perturbatively. A series in $s$ is then obtained ${\cal P}%
_{s}^{>}=1+\sum_{n=1}^{\infty }c_{n}(y)s^{n}$ with the $c_{n}$'s having an
expression similar to (\ref{co9}). The $n$-th moment $\epsilon ^{n}$
averaged over $\sigma _{>}$ coincides with $c_{n}(1/{\rm Pe})$, up to simple
combinatorial factors. The final result is 
\begin{equation}
\langle \epsilon ^{n}\rangle _{>}=(2n-1)!!\,\,n!\,\left( \frac{\beta }{%
\alpha }\right) ^{n-1/2}\frac{a_{n}\sqrt{\beta }}{\mbox{Pe}},  \label{e1}
\end{equation}
where the coefficients $a_{n}$ are given by 
\begin{equation}
a_{n}=\int\limits_{0}^{\infty
}dy_{2}G[y_{2}^{2}]\int\limits_{0}^{y_{2}}dy_{3}\int\limits_{y_{3}}^{\infty
}dy_{4}G[y_{4}^{2}]\cdots
\int\limits_{0}^{y_{2n-2}}dy_{2n-1}\int\limits_{y_{2n-1}}^{\infty
}dy_{2n}G[y_{2n}^{2}].  \label{P>12}
\end{equation}
Equation (\ref{e1}), together with (\ref{ig12}) allowing to calculate small
times averages, gives the expression of all integer moments of the
dissipation field for a general form of pumping. It follows from (\ref{e1})
that all these moments scale with the same power of ${\rm Pe}$. This shows
that the strong intermittency evidenced in the analysis of scalar
differences in the convective range comes down into the dissipative range.
The analysis of the $n$-dependence of the constants in (\ref{e1}) actually
shows that the intermittency of the dissipation field is even stronger than
for scalar differences. To this aim, it is convenient to restrict to a
particular form of pumping, allowing to proceed with explicit calculations.
Specifically, let us consider $G^{*}[x^{2}]=\exp [-2x]$. The correlation
function of the corresponding pumping has Fourier transform $\chi _{q}^{*}=4%
\sqrt{\pi }/q^{4}\exp \left( -1/q^{2}\right) $. Equation (\ref{P>4}) with $G$
having the specific form $G^{*}$ is solved in terms of the Bessel function $%
I_{0}$ as 
\begin{equation}
{\cal P}_{s}^{>}=\frac{I_{0}\left( \sqrt{s\beta /\alpha }e^{- \sqrt{\alpha }/%
\mbox{Pe}}\right) }{I_{0}(\sqrt{s\beta /\alpha })}\rightarrow \Biggl|_{%
\mbox{Pe}\gg 1}1-\frac{\sqrt{s\beta }}{\mbox{Pe}}\frac{I_{1}\left( \sqrt{%
s\beta /\alpha }\right) }{I_{0}(\sqrt{s\beta /\alpha })}.  \label{e05}
\end{equation}
The advantage with respect to the general case with arbitrary form of
pumping is clearly that ${\cal P}_{s}^{>}$ is now known explicitly and this
will permit us to reconstruct the PDF of the dissipation field.

Having averaged over $\sigma _{>}$, we need now to take into account
fluctuations at small times, i.e. average over $\sigma _{<}$. The important
remark is that, both in the general case (\ref{e1}) and in (\ref{e05}), we
need to average quantities of the form $\sqrt{\beta }f(\alpha /\beta )$,
with $f$ arbitrary but having the property that it depends only on $\alpha
/\beta $. For our purposes, it is then appropriate to introduce the random
variable $\mu =\alpha /\beta $ and consider its distribution function
weighted with $\sqrt{\beta }$\thinspace : 
\begin{equation}
{\cal P}^{<}[\mu ]\equiv {\frac{\int {\cal D}\sigma _{<}\exp \left(
-S_{<}\right) \,\,\delta \left( \mu -{\frac{\alpha }{\beta }}\left( \{\sigma
_{<}\}\right) \right) \sqrt{\beta }\left( \{\sigma _{<}\}\right) }{\int 
{\cal D}\sigma _{<}\exp \left( -S_{<}\right) }}.  \label{ig1}
\end{equation}
It is again convenient for averaging to introduce the Laplace transform of
the PDF ${\cal P}_{s}^{<}=\langle \exp \left( -s\mu \right) \rangle _{<}$.
As in (\ref{P>1}), its expression can be presented as the following quantum
mechanical matrix element\thinspace : 
\begin{equation}
{\cal P}_{s}^{<}=\int\limits_{0}^{\infty }e^{-s\mu }{\cal P}^{<}[\mu ]d\mu
=e^{-t_{0}/4}\Biggl\langle\delta (\eta )\Biggl| \exp \left[ -t_{0}\hat{H}%
\right] \Biggr|e^{\eta /2}\Biggr\rangle=\left[ e^{-t_{0}/4}\Psi (t_{0};\eta
)\right] _{\eta =0}.  \label{ig4}
\end{equation}
Here, $\Psi (t;\eta )$ satisfies $(\partial _{t}-\hat{H})\Psi =0$, the
Hamiltonian is $\hat{H}(\eta ;s)=-\partial _{\eta }^{2}+s\exp \left( -2\eta
\right) $, the space variable $\eta =\int_{0}^{t}\sigma $ and the initial
condition for the ``wave function'' $\Psi (t;\eta )$ is $\exp \left( \eta
/2\right) $. The asymptotic behavior at large times $t$ can be obtained as
in (\ref{P>4}), noting that the potential part in $\hat{H}$ decreases at
infinity and $\Psi (0;\eta )$ does not. It follows 
\begin{equation}
\Psi (t;\eta )\to \biggl|_{t\to \infty }e^{t/4}\Psi _{0}(\eta )=e^{t/4}\exp
\left[ \eta /2-\sqrt{s}\exp [-\eta ]\right] ,  \label{ig7}
\end{equation}
where $\Psi _{0}(\eta )$ satisfies $\left( \hat{H}+1/4\right) \Psi _{0}=0$
and behaves as $\exp \left( \eta /2\right) $ for large $\eta $'s. Requiring $%
\exp \left( t_{0}/4\right) \gg 1$, we can plug the asymptotic expression (%
\ref{ig7}) into (\ref{ig4}) and obtain ${\cal P}_{s}^{<}=\exp \left[ -\sqrt{s%
}\right] $. The PDF of $\mu $ and the moments relevant for $\epsilon ^{n}$
are easily restored as 
\begin{eqnarray}
&&{\cal P}^{<}(\mu )=\frac{1}{2\pi i}\int\limits_{0^{+}-i\infty
}^{0^{+}+i\infty }ds\,{\cal P}_{s}^{<}e^{s\mu }=\frac{1}{2\pi i}%
\int\limits_{0^{+}-i\infty }^{0^{+}+i\infty }ds\,e^{- \sqrt{s}}\,e^{s\mu }=%
\frac{1}{2\sqrt{\pi }\mu ^{3/2}}\exp \left[ -\frac{1}{4\mu }\right] ,
\label{ig11} \\
&&\Biggl\langle \left( {\frac{\beta }{\alpha }}\right) ^{n- 1/2}\sqrt{\beta }%
\Biggl\rangle_{<}=\int\limits_{0}^{\infty }d\mu \mu ^{- n+1/2}{\cal P}%
^{<}(\mu )=2^{2n-1}\frac{(n-1)!}{\sqrt{\pi }}.  \label{ig12}
\end{eqnarray}
This expression can be used to calculate the moments appearing in (\ref{e1}%
). Note that the expression (\ref{e1}), derived exploiting the time scales
separation, coincides for the first moment $\langle \epsilon \rangle $ with
the dominant order of the known solution derived in Section~\ref{ss:model}%
\thinspace : 
\begin{equation}
\langle \epsilon \rangle =\frac{1}{\pi \mbox{Pe}}\int\limits_{0}^{\infty }%
\frac{\left[ \chi (0)-\chi (x)\right] dx}{x^{2}+\mbox{Pe}^{-2}}.  \label{e}
\end{equation}

One step further can be made for the specific form of pumping $%
G^{*}[x^{2}]=\exp [-2x]$, allowing to obtain the explicit solution (\ref{e05}%
). From (\ref{e05}), we obtain indeed the Laplace transform of the PDF for $%
Q $ (averaged over both $\sigma_{<}$ and $\sigma_{>}$) as 
\begin{equation}
{\cal P}_{s}\equiv \langle {\cal P}_{s}^{>}\rangle _{<}\to 1- \frac{\sqrt{s}%
}{\mbox{Pe}}\int\limits_{0}^{\infty }d\mu {\cal P}^{<}(\mu )\frac{I_{1}(%
\sqrt{s/\mu })}{I_{0}(\sqrt{s/\mu })}=1-\frac{1}{\sqrt{\pi }\mbox{Pe}}%
\int\limits_{0}^{\infty }\ln [I_{0}(2\sqrt{xs})]e^{-x}dx.  \label{s1}
\end{equation}
The moments $\langle\epsilon^n\rangle$ can be immediately reconstructed,
according to (\ref{06}), from the derivatives of (\ref{s1}) at $s=0$. They
read 
\begin{equation}
\langle \epsilon ^{n}\rangle =\frac{(2n-1)!!n!}{\sqrt{\pi }\mbox{Pe}}\frac{%
\partial ^{n}}{\partial z^{n}}\ln \left[ \frac{1}{J_{0}(\sqrt{z})}\right] %
\Biggl|_{z=0}=i\frac{\Gamma (n+1/2)[\Gamma (n+1)]^{2}2^{n}}{2\pi ^{2}%
\mbox{Pe}}\int\limits_{C}\frac{\ln [J_{0}(\sqrt{z})]dz}{z^{n+1}},  \label{en}
\end{equation}
where $C$ is a close contour about $z=0$ in the complex $z$- plane. The
Cauchy integral representation (\ref{en}) is useful for getting the large $n$
asymptotic of (\ref{en}). The integral is indeed saddle around the square of
the first zero of the Bessel function $z_{0}=x_{0}^{2}$, $J_{0}(x_{0})=0$.
The saddle estimation for the integral is $\sim x_{0}^{n}$, thus giving 
\begin{equation}
\ln [\mbox{Pe}\langle \epsilon ^{n}\rangle ]\to \Biggl|_{n\to \infty
}n\left( 3\ln [n]-3+\ln [2x_{0}]\right) .  \label{en1}
\end{equation}

The whole expression (\ref{s1}) can actually be inverted. From (\ref{06}),
it follows indeed that the PDF ${\cal P}^{(\epsilon )}(\epsilon )$ of the
dissipation $\epsilon $ is given by 
\begin{eqnarray}
{\cal P}^{(\epsilon )}(\epsilon ) &=&\frac{1}{2\sqrt{\pi }}\frac{1}{2\pi i}%
\int\limits_{0}^{\infty }dx\frac{e^{- x}}{x^{3/2}}\int\limits_{0^{+}-i\infty
}^{0^{+}+i\infty }ds{\cal P}_{s}\exp \left[ s\epsilon /(2x)\right]  \nonumber
\\
&=&\frac{1}{\mbox{Pe}}\int\limits_{0}^{\infty }dz\ln \left[ \frac{1}{I_{0}(z)%
}\right] \left[ \frac{1}{\sqrt{2}\pi }\frac{\mbox{$_0F_2$}\left(
1/2,1/2;\epsilon z^{2}/8\right) }{\sqrt{\epsilon }}- \frac{z}{\sqrt{\pi }}%
\mbox{$_0F_2$}\left( 1,3/2;\epsilon z^{2}/8\right) \right] ,  \label{pdf}
\end{eqnarray}
where $_{q}F_{p}$ is the generalized hypergeometric function with the $q$
parameters in the numerator and the $p$ parameters in the denominator. The $%
\delta $-function at the origin arising from the unit term in (\ref{e05})
has not been considered in (\ref{pdf}). The reason is that, as we shall see
in a moment, the range of validity of (\ref{pdf}) is $\epsilon \gg \mbox{Pe}%
^{-2}$.

It is indeed time to clarify the limits of validity of the calculations
performed. To be concrete, we shall specifically consider the pumping $G^{*}$
leading to (\ref{e05}). The remainder of the expansion over $\mbox{Pe}^{-1}$
in (\ref{e05}) is bounded by ${\beta }\left( I_{1}\left( \sqrt{s\beta
/\alpha }\right) \sqrt{{\frac{s\alpha }{\beta }}}+sI_{1}^{\prime }\left( 
\sqrt{s\beta /\alpha }\right) \right) /(2\mbox{Pe}^{2})$. Contrary to (\ref
{ig12}), we need now to consider averages of $\mu =\alpha /\beta $ with
weight $\beta $. The problem can again be reduced to the calculation of a
quantum mechanical matrix element and it is found that the final result
depends on $t_0$ as $\exp (2t_{0})$. The condition for the remainder to be
subdominant with respect to the terms kept in (\ref{e05}) is therefore that $%
\exp (2t_{0})/\mbox{Pe}$ should vanish as $\mbox{Pe}\to \infty $. On the
other hand, the observation time must be much larger than unity in order to
attain the stationary state, i.e. $t_0\gg 1$. This condition is clearly
compatible with the previous one, in the limit of large P\'eclet numbers,
and gives the ordering $1\ll t_{0}\ll \ln [\mbox{Pe}]$. The other delicate
point is the criterium of applicability of (\ref{s1}) with respect to $s$.
For the expansion in (\ref{e05}) to be meaningful, the second term should be
much smaller than unity. This shows that we should require $s\ll \mbox{Pe}%
^{2}$. The expansion (\ref{e05}) fails then to describe the large $s$ tails
of ${\cal P}_s$ and thus the smallest values of $\epsilon $ in the
respective PDF (the relation between large $s$ and small $\epsilon $ is
direct since the decay of the generating function is relatively slow).

Another simple approximation is however available for the high $s$ limit.
The trajectories contributing to $\langle \exp \left( -sQ\right) \rangle $
at large $s\gg {\rm Pe}^{2}$ are clearly those where $Q$ is small. From the
definition (\ref{07}) it follows that, for this to happen, $W^{2}(t)$ should
remain small all the time. The quantity $W^{2}\int^{t}W^{-2}$ is in this
case $O(1)$ and the argument in $G$ is small on account of the $1/{\rm Pe}%
^{2}$ factor. For the trajectories relevant at high $s$, it is thus possible
to approximate $Q$ by $G[0]/\mbox{Pe}^{2}\int W^{2}$. The Laplace transform $%
{\cal P}_{s}$ reduces then to 
\begin{equation}
{\cal P}_{s}\equiv \Biggl\langle \exp \left[ -\frac{sG[0]}{\mbox{Pe}^{2}}%
\int\limits_{0}^{T}dt\,W^{2}(t)\right] \Biggr\rangle =e^{-T/4}\Biggl\langle%
\delta (\eta )\Biggl| \exp \left[ -T\hat{H}(-\eta ;sG[0]/\mbox{Pe}%
^{2})\right] \Biggr|e^{-\eta /2}\Biggr\rangle,  \label{Ps1}
\end{equation}
where $\eta (t)\equiv \int\limits_{0}^{t}\sigma (t)$ and $\hat{H}$ is the
same as for (\ref{ig4}). The matrix element (\ref{Ps1}) actually coincides
identically with (\ref{ig4}) when $t_{0}$ by $T$, $\eta $ by $-\eta $, and $s
$ by $sG[0]/\mbox{Pe}^{2}$ are replaced. Using (\ref{ig11}), one can then
easily get the final answer 
\begin{equation}
\mbox{at}\quad s\gg \mbox{Pe}^{2},\quad {\cal P}_{s}\to \exp \left[ -\frac{%
\sqrt{sG[0]}}{\mbox{Pe}}\right] .  \label{Ps2}
\end{equation}
Inverting (\ref{Ps2}), we can obtain the PDF of $\epsilon $ at the small
values $\epsilon \ll {\rm Pe}^{-2}$. For larger values of $\epsilon $, the
PDF follows from the general formula (\ref{pdf}). The following general
behavior for the dissipation field PDF is thus obtained 
\begin{equation}
{\cal P}^{(\epsilon )}(\epsilon )\rightarrow \cases{
\frac{\sqrt{2}\mbox{Pe}}{\pi \sqrt{G[0]}}\frac{1}{\sqrt{\epsilon }}\left(
1-\frac{2\mbox{Pe}^{2}}{G[0]}\epsilon \right),&for $\epsilon\ll {\rm
Pe}^{-2}$;\cr \cr \frac{1}{2\sqrt{2}\pi
}\frac{1}{\mbox{Pe}}\frac{1}{\epsilon ^{3/2}},&for $\mbox{Pe}^{- 2}\ll
\epsilon \ll 1$;\cr \cr \frac{\exp \left[ -(\epsilon /\epsilon
_{0})^{1/3}\right] }{\mbox{Pe}},\quad \epsilon _{0}\sim 1,&for $\epsilon \gg
1$,\cr}  \label{pdfs}
\end{equation}
where algebraic prefactors have not been considered in region of exponential
fall off. The PDF for scalar gradients $\omega =\sqrt{\epsilon /\kappa }$
follows immediately from (\ref{pdfs}) as 
\begin{equation}
{\cal P}^{\omega }(\omega )\rightarrow \cases{ {L\over \pi}\left(1-{L^2\over
G[0]}\omega^2\right),&for $\omega\ll 1/L$;\cr \cr {1\over 2\pi L}{1\over
\omega^2},&for $1/L\ll \omega\ll 1/\sqrt{\kappa}$;\cr \cr
\frac{\kappa}{{\mbox{Pe}}} \exp \left[ -(|\omega |\sqrt{\kappa /\epsilon
_{0}})^{2/3}\right],&for $\omega\gg 1/\sqrt{\kappa}$.\cr}  \label{pdfs1}
\end{equation}

Moments of the dissipation $\langle \epsilon ^{a}\rangle $, with $a>1/2$,
are all proportional to $1/{\rm Pe}$, in agreement with (\ref{e1}) valid for
arbitrary pumping. Moreover, we learn from (\ref{pdfs}) that the linear
scaling found for scalar differences low-order moments comes down to the
dissipative range. Moments $\langle \epsilon ^{a}\rangle $ with $-1/2<a\le
1/2$ scale indeed as $\mbox{Pe}^{-2a}$. Moments with $a<-1/2$ do not exist.
>From (\ref{pdfs1}), it also follows that the same tendency observed for
scalar differences PDF to develop a cusped structure at the origin is
present. An important difference arises for the tails. The comparison
between (\ref{pdfs1}) and (\ref{sd11a}) for gradients and scalar differences
indicate indeed that the former decrease much more slowly. Such behavior is
physically understood in terms of fluctuations of the dissipative scale as
follows. Gradients can be thought of as scalar differences evaluated at the
dissipative scale. This scale is however a dynamical quantity and fluctuates
(strongly in a problem with inverse transfer as the one here). The
statistics of these fluctuations are actually related precisely to those of
the variable $\mu $, introduced in (\ref{ig1}). The fluctuations of $%
\epsilon ^{n}$ are therefore the product of those of scalar differences {\em %
and} those of the dissipative scale. The latter do not depend on the
dimensional parameters of the problem, i.e. they are universal as follows
from (\ref{ig11}), but grow factorially with the order $n$. This factorial
is precisely what shifts the dominant term in (\ref{en1}) from $2n\ln n$ to $%
3n\ln n$. It follows then that the slower decay of the tails for gradients
with respect to those for scalar differences is indeed due to dynamical
fluctuations of the dissipative scale.

\section{Conclusions and discussion}

We analyzed in the present paper statistics of the passive scalar advected
by a one-dimensional, smooth and fast-correlated in time velocity field. In
spite of its extreme simplicity, the dynamics presents very interesting and
surprising behaviors. We expect that many of them are quite generic and it
may be interesting to look for more realistic scalar (and generally
turbulent) problems where the new effects discussed here could be of
relevance. Thus, hereafter we recall and briefly comment the major
properties of the model discussed in the paper and address, in parallel, the
respective questions for future studies.

The first major property of the dynamics is {\em the inverse cascade of the
scalar}. The inverse cascade is a consequence of compressibility, but in a
subtle way\thinspace : in \cite{97CKVb} we consider indeed a generalized
smooth $d-$dimensional model having the degree of compressibility as a free
parameter. A transition between inverse and direct cascades is observed
there\thinspace : if $d>4$, the cascade is always direct, independently of
the degree of compressibility\thinspace ; if the latter is small enough, the
cascade is direct again\thinspace ; otherwise, it is inverse. It might also
be interesting to look from the same dynamical point of view used here at
the critical dimension appearing in the passive vector problem discussed in 
\cite{96Mas} and at the role of compressibility for the direction of energy
transfer in other turbulent situations, e.g. 2D Navier-Stokes and MHD
turbulence.

Dynamically, the inverse cascade is associated with the fact that the
Lyapunov exponent for Lagrangian trajectories is negative. This implies
that, {\em in the case of inverse cascade, contraction of Lagrangian
trajectories is typical and stretchings are relatively rare events}. This is
precisely the opposite of the canonical picture associated with the direct
cascade. Rare trajectories (contracting or stretching if the Lyapunov
exponent is positive or negative, respectively) are responsible for the
intermittent part of the problem. This definetely confirms the picture that
contractive trajectories play a crucial role for intermittency in
incompressible turbulent transport.

Negative Lyapunov exponent leads to the third observation emerging from the
analytical study of the model\thinspace : {\em Gaussianity of the scalar
establishes at scales larger than the scale of the pumping, while strong
intermittency is present at small scales}. Small-scales intermittency found
here is of the ``Burgers'' kind, i.e. all scalar differences integer moments
scale linearly $\sim r/L$. This result gives a definite ground to speculate
that, generally, the stronger is the degree of compressibility (the role of
contracting trajectories is then enhanced), the more intermittency
downscales of the pumping is observed. Note in the respect of Gaussianity
that the inverse cascade of energy in 2D Navier-Stokes turbulence is also
Gaussian according to the numerics \cite{93SY}. It rises yet another
conjecture of a certain generality of the Gaussianity feature for a great
variety of direct cascades.

The equation for scalar differences PDF, derived directly from the dynamics
in Section \ref{s:differences}, shows that {\em no dissipative anomaly is
present} in the model. In the generalized one-dimensional model introduced
in \cite{97VM}, $\gamma =1$ is a natural threshold for the inverse cascade.
At $\gamma >1$ the cascade is direct, while it is inverse at $\gamma <1$.
Most probably, it means that there is no dissipative anomaly (equation for
the PDF of the scalar difference could be closed) for $\gamma <1$. However,
it is not yet clear how to derive this consistently (as it is derived here
for the smooth limit $\gamma =0$). Note that the general question of the
role of the dissipative anomaly in the passive scalar physics is not yet
resolved in any sense (some suggestions on this point may be found in \cite
{94Kr,97Yaka,97Yakb}). One more unresolved question is how does the
dissipative anomaly (if any) affects the anomalous scaling behavior of
structure functions of integer (and generally all the) orders.

An important result for the issue of convection-diffusion interplay, is the
fact that {\em the effective dissipative scale}, which sets the crossover
scale between scalar differences $2n$-th moment in the convective range and
dissipation field $n$-th moment, {\em is a strongly fluctuating quantity},
growing factorially with $n$. To make this statement, we have analytically
calculated in Section \ref{s:dissipation} the dissipation field PDF
exploiting a new scale-separation procedure, which is also used in Appendix 
\ref{s:Longtime} to describe the long-time dynamics of the pair correlation
function of the scalar gradients. The procedure is likely to be a general
tool for resolving the problem in other situations. Particularly, a slight
(matrix) modification of the method should be relevant for the Kraichnan $d-$%
dimensional incompressible smooth model.

Finally, the eddy-diffusivity operator is Hermitian for Kraichnan model,
where random velocity is incompressible and fast. ${\em The}$ {\em %
Hermiticity is lost, when compressible flow is considered}. Generally, it
might be interesting to look at scalar transport (and particularly at flow
with finite correlation times) from the point of view of non-Hermiticity.
This issue is the subject of a very recent interest in the field of
disordered systems \cite{96HN,97Efe,97FKS,97CW}. The specific object of
interest for scalar transport is the resolvent ${\cal R}(t;x,y)$, describing
the probability to find a Lagrangian separation equal to $x$ at time $t$,
having initially been equal to $y$. One may then want to study the
distribution in the complex plane of the poles of its Laplace transform. A
width of the poles domain along the imaginary direction might be a
characteristic of the trapping degree.

\vskip 0.3cm \centerline{\bf Acknowledgments} \vskip 0.15cm

Many stimulating discussions with E.~Balkovsky, G.~Falkovich, U.~Frisch,
K.~Gawedzki, V.~Lebedev, A.~Polyakov, B.~Shraiman, V.~Yakhot are gratefully
acknowledged. We are also grateful to organizers and participants of the
workshop on turbulence in IHES, Bures-sur-Yvette, where part of this work
was done. This research project was partly supported by a R.H.~Dicke
fellowship and ONR/DARPA URI Grant N00014-92-J-1796 (MC), Russian Fund of
Fundamental Researches under grand 97-02-18483 (IK), and the GdR
``M\'{e}canique des Fluides G\'{e}ophysiques et Astrophysiques'' (MV).

\appendix

\section{Dynamical formulation by field formalism}

\label{s:DynForm}

Let us consider the gradient field $\omega (t,x)$ satisfying the equation of
motion (\ref{omega}) with the velocity $u(t,x)$ having Gaussian statistics
and correlation function $<u(t,x)$ $u(t^{\prime },x^{\prime
})>=V(t-t^{\prime },x-x^{\prime })$. The generating functional ${\cal F}%
\left( \lambda \right) $ for its simultaneous correlation functions can be
written in the form of functional integral (see \cite{76Dom,76Jan}): 
\begin{equation}
{\cal F}\left( \lambda \right) =\int {\cal D}p{\cal D}\omega {\cal D}u\,\exp
\left\{ \frac{1}{2}\left( p|\chi ^{\prime \prime }|p\right) - \frac{1}{2}%
\left( u|V^{-1}|u\right) +i\int\limits_{0}^{T}dtdx\,p\left( \partial
_{t}\omega +\partial _{x}(u\omega )-\kappa \partial _{x}^{2}\omega \right)
+\int dx\lambda \left( x\right) \omega \left( T,x\right) \right\}
\label{fint}
\end{equation}
Here, $\left( u|V^{-1}|u\right) $ denotes the diagonal matrix element of the
inverse operator to $V(t-t^{\prime },x-x^{\prime })$ in the Hilbert space of
the functions $u\left( t,x\right) $ and the pumping $\chi ^{\prime \prime }$
is the second spatial derivative of the forcing correlation function $\chi $%
, defined in (\ref{forza}). The retarded regularization of the time
derivative, $\left( \partial _{t}\omega \right) _{n}=\frac{1}{\Delta }\left(
\omega _{n}-\omega _{n-1}\right) ,\,\,$does not produce any nontrivial
Jacobian after derivation of (\ref{fint}) from the equation of motion. $%
\Delta \rightarrow 0$ is the temporal slicing and $\omega _{n}\left(
x\right) \equiv \omega \left( n\Delta ,x\right) $. Performing the Gaussian
integration over the field $u\left( t,x\right) $ we arrive at 
\begin{equation}
{\cal F}\left( \lambda \right) =\int {\cal D}p{\cal D}\omega \,\exp \left\{ -%
\frac{1}{2}\left( \omega \partial _{x}p|V|\omega \partial _{x}p\right) +%
\frac{1}{2}\left( p|\chi ^{\prime \prime }|p\right)
+i\int\limits_{0}^{T}dtdx\,p\left( \partial _{t}\omega -\kappa \partial
_{x}^{2}\omega \right) +\int dx\lambda \left( x\right) \omega \left(
T,x\right) \right\} .  \label{iig1}
\end{equation}
For the case of $\delta $-correlated in time velocity fields considered in
this paper, we replace the general $V(t-t^{\prime },x-x^{\prime })$ by
specific $\delta \left( t-t^{\prime }\right) V(x-x^{\prime })$. The explicit
version of (\ref{fint}) reads then as 
\begin{eqnarray}
&&{\cal F}\left( \lambda \right) =\int {\cal D}p{\cal D}\omega \,\exp \left(
-{\cal S}+\int dx\lambda \left( x\right) \omega \left( T,x\right) \right)
\label{ac} \\
{\cal S} &=&\frac{1}{2}\int\limits_{0}^{T}dt\left[ \int dxdx^{\prime
}(\omega \partial _{x}p)\left( t,x\right) V\left( x-x^{\prime }\right)
(\omega \partial _{x^{\prime }}p)\left( t,x^{\prime }\right) - \int
dxdx^{\prime }p\left( t,x\right) \chi ^{\prime \prime }\left( {\frac{%
x-x^{\prime }}{L}}\right) p\left( t,x^{\prime }\right) \right]  \nonumber \\
&&-i\int\limits_{0}^{T}dtdx\,p\left( \partial _{t}\omega -\kappa \partial
_{x}^{2}\omega \right) .  \label{act}
\end{eqnarray}

Despite of the explicit presence of the $V$-term in (\ref{act}), the
dynamical part of the action $S$ is covariant with respect to Galilean
transformations. This leads to the important Ward identity: 
\begin{equation}
\int {\cal D}p{\cal D}\omega \,\exp \left( -{\cal S}+\int dx\lambda \left(
x\right) \omega \left( T,x\right) \right) \prod\limits_{j}\left( \int
dx\omega (t_{j},x)\partial _{x}p(t_{j},x)\right) =0,  \label{Ward1}
\end{equation}
where set $\left\{ t_{j}<T,j=1,2,...\right\} $ is arbitrary. To prove it,
let us consider the change of variables 
\begin{equation}
p\left( t,x\right) \longrightarrow p\left( x+\int\limits_{t}^{T}v(\tau
)d\tau ,t\right) ,\,\,\omega \left( t,x\right) \longrightarrow \omega \left(
x+\int\limits_{t}^{T}v(\tau )d\tau ,t\right)  \label{trans}
\end{equation}
in the functional integral (\ref{ac}) with $v\left( \tau \right) $ being
unconstrained. The transformation does not change the source term and it has
unit Jacobian. The variation $\delta {\cal S}$ of the action ${\cal S}$
under the transformation reads as 
\begin{equation}
\delta {\cal S}=-i\int\limits_{0}^{T}dt\,v(t)\left( \int dx\,\omega \left(
t,x\right) \partial _{x}p(t,x)\right) .  \label{var}
\end{equation}
The change of variables does not change the value of the integral. Thus, all
the functional derivatives of (\ref{ac}) over $v(t)$ are equal to zero.
Taking into account (\ref{var}), one finally arrives at the Ward identity (%
\ref{Ward1}).

All written above is applicable for any short correlated velocity field. Now
let us narrow consideration to the case of smooth velocity field (\ref{corr}%
) with $\gamma =0$. One gets 
\begin{eqnarray}
{\cal F}\left( \lambda \right) &=&\int {\cal D}p{\cal D}\omega \exp \left( -%
{\cal S}+\int dx\lambda \left( x\right) \omega \left( T,x\right) \right) ,
\label{act0} \\
{\cal S} &=&\frac{1}{2}\int\limits_{0}^{T}dt\left[ -\int dxdx^{\prime
}(\omega \partial _{x}p)\left( t,x\right) \left( x-x^{\prime }\right)
^{2}(\omega \partial _{x^{\prime }}p)\left( t,x^{\prime }\right) -\int
dxdx^{\prime }p\left( t,x\right) \chi ^{\prime \prime }\left( {\frac{%
x-x^{\prime }}{L}}\right) p\left( t,x^{\prime }\right) \right] \\
&&-i\int\limits_{0}^{T}dtdx\,p\left( \partial _{t}\omega -\kappa \partial
_{x}^{2}\omega \right) ,
\end{eqnarray}
where the $V_{0}$ term was dropped due to (\ref{Ward1}). The goal is to
reduce the problem to averaging of a functional of one random in time scalar
field, analogously to the random matrix description used in two- dimensional 
\cite{95CFKLa,95BCKL} and generally $d$-dimensional case. However, in the
present case, compressibility calls for more evaluations. Let us perform the
following change of variables in (\ref{act0}): 
\begin{equation}
\left\{ p\left( t,x\right) ,\,\omega \left( t,x\right) \right\} \rightarrow
\left\{ p\left( t,x+i\int\limits_{t}^{T}d\tau \int dy\,\,y^{2}\partial
_{y}p(t,y)\omega (t,y)\right) ,\,\omega \left( t,x+i\int\limits_{t}^{T}d\tau
\int dy\,\,y^{2}\partial _{y}p(t,y)\omega (t,y)\right) \right\} ,
\label{iig3}
\end{equation}
which can be considered as the field-dependent and time-dependent
homogeneous spatial translation leaving the source term intact. Analytic
continuation of the fields which makes the spatial arguments to be real is
assumed. Action gets the following from being rewritten in the new variables 
\begin{equation}
{\cal S}=\frac{1}{2}\int\limits_{0}^{T}dt\left[ \left( \int dx\,x(\omega
\partial _{x}p)\left( t,x\right) \right) ^{2}-\int dxdx^{\prime }p\left(
t,x\right) \chi ^{\prime \prime }\left( {\frac{x-x^{\prime }}{L}}\right)
p\left( t,x^{\prime }\right) \right] -i\int\limits_{0}^{T}dtdx\,p\left(
\partial _{t}\omega -\kappa \partial _{x}^{2}\omega \right) .  \label{act3}
\end{equation}
The transformation (\ref{iig3}) has non-trivial Jacobian 
\begin{equation}
{\cal D}p{\cal D}\omega \,\longrightarrow {\cal D}p{\cal D}\omega \,{\cal J}%
\left( p,\omega \right) ,  \label{jac1}
\end{equation}
which depends on regularization. The regularization is fixed by the
requirement for the temporal $\delta -$ function from the correlation
function of velocities to appear as a result of narrowing of an even
function of temporal argument. Besides, the correctly regularized action (%
\ref{act3}) should reproduce the respective correlation function in the
limit $T\rightarrow 0$: 
\begin{equation}
p_{n}\left( x\right) \longrightarrow p_{n}\left( x+\frac{i}{2}\Delta \int
dy\,\,y^{2}\partial _{y}p_{n}\left( y\right) \omega _{n}\left( y\right)
+i\Delta \sum\limits_{m=n+1}^{N}\int dy\,\,y^{2}\partial _{y}p_{m}\left(
y\right) \omega _{m}\left( y\right) \right) ,  \label{regg}
\end{equation}
respectively for $\omega _{n}\left( x\right) $. The Jacobian ${\cal J}\left(
p,\omega \right) $can be computed easily 
\begin{equation}
{\cal J}\left( p,\omega \right) =\exp \left( -i\int dtdx\,x\omega \partial
_{x}p\right) .  \label{jac2}
\end{equation}
The exponential of the term $\int\limits_{0}^{T}dt\left( \int dx\,x\omega
\partial _{x}p\right) ^{2}$ in the action (\ref{act3}) multiplied by the
Jacobian (\ref{jac2}) can be represented by means of averaging of the
exponential of $i\int\limits_{0}^{T}dt\sigma \int dxx\omega \partial _{x}p$
over the auxiliary field $\sigma (t)$. The Gaussian measure of averaging
over $\sigma $ is defined in (\ref{q6b}). It results in the following
representation for the generating functional ${\cal F}\left( \lambda \right) 
$: 
\begin{eqnarray}
&&{\cal F}\left( \lambda \right) =\int {\cal D}p{\cal D}\omega {\cal D}f\,%
{\cal D}\sigma \exp \left[ -{\cal S}_{\sigma }-\frac{1}{2}\left( f|\chi
^{-1}|f\right) \right]  \label{nonumber} \\
\times &&\exp \left[ i\int\limits_{0}^{T}dtdx\,p\left( \partial _{t}\omega
-\sigma \partial _{x}(x\omega )-\kappa \partial _{x}^{2}\omega -\partial
_{x}f\right) +\int dx\lambda \left( x\right) \omega \left( T,x\right)
\right] .  \label{dynact}
\end{eqnarray}
Integration over $p\left( t,x\right) $ gives the reduced equation of motion: 
\begin{equation}
\partial _{t}\omega -\sigma \partial _{x}(x\omega )-\kappa \partial
_{x}^{2}\omega =\partial _{x}f,  \label{ur}
\end{equation}
averaged with respect to $f\left( t,x\right) $ and $\sigma \left( t\right) $
with the weights given by the first line of (\ref{dynact}). It should be
stressed that this equivalence holds only at the level of simultaneous
correlation functions.

\section{Long-time behavior of the pair correlation function of gradients.}

\label{s:Longtime}

In the present Appendix, we will find the long-time asymptotic of the pair
correlation function of the scalar gradients $\Omega \left( t,x\right) $
which obeys (\ref{boh}). We will model here the infrared cut-off of the
velocity correlations by

\begin{equation}
S\left( x\right) =x^{2},\,x<L_{u},\;\;S\left( x\right) =L_{u}^{2},\,x>L_{u}.
\label{str}
\end{equation}
The Laplace transform of (\ref{boh}) has the form

\begin{equation}
s\Omega _{s}-\partial _{x}^{2}\left( S\left( x\right) +2\kappa \right)
\Omega _{s}+\chi ^{\prime \prime }\left( x\right) /s=0.  \label{lapur}
\end{equation}
The solution of (\ref{lapur}) 
can be written in terms of the Green function $%
{\cal G}_{s}\left( x,x^{\prime }\right) $ (which is taken to be even in $x$, $%
{\cal G}_{s}\left( x,-x^{\prime }\right) ={\cal G}_{s}\left( x,x^{\prime
}\right) $) as
\begin{equation}
\Omega _{s}\left( x\right) =-\frac{1}{s}\int\limits_{0}^{\infty }{\cal G}%
_{s}\left( x,x^{\prime }\right) \chi ^{\prime \prime }\left( x^{^{\prime
}}\right) dx^{^{\prime }},\hspace{0.5cm}s{\cal G}_{s}-\partial
_{x}^{2}\left( S\left( x\right) +2\kappa \right) {\cal G}_{s}=\delta \left(
x-x^{^{\prime }}\right) .  \label{FG}
\end{equation}
The desired long-time asymptotic corresponds to the smallest $s$ and our
aim is to find the singular
structure of $\Omega _{s}\left( x\right) $ at small $s$ and $x\gg \sqrt{%
\kappa }$ . In what
follows, we will then neglect all the logarithmic contributions like $s\ln
L_{u}$ and $s\ln x$ in comparison with $1$.
We will also neglect the smallest $x^{\prime }$ ($x^{\prime }\ll 
\sqrt{\kappa }$) contribution into the integral (\ref{FG}) going to zero in
the limit of small diffusivity. Therefore ${\cal G}_{s}\left( x,x^{\prime
}\right) $ should be studied at $\sqrt{\kappa }\ll $ $x^{\prime }\ll L_{u}$
and in all the allowed domains with respect to $x$.

Making the substitution ${\cal G}_{s}\left( x,x^{\prime }\right) =g\left(
x,x^{\prime }\right) /[S\left( x\right) +2\kappa ]$, one gets

\begin{equation}
\frac{s}{S\left( x\right) +2\kappa }g-\partial _{x}^{2}g=\delta \left(
x-x^{^{\prime }}\right) .  \label{smur}
\end{equation}
Solution of (\ref{smur}) in the domain of the largest $x$, $%
x>L_{u}>x^{\prime }\gg \sqrt{\kappa }$, can thus be written as

\begin{equation}
g\left( x,x^{\prime }\right) =A\exp \left( -\frac{\sqrt{s}}{L_{u}}\left(
x-L_{u}\right) \right) .  \label{great}
\end{equation}
Note however that only the first two terms of the (\ref{great}) expansion in 
$s$ will be required for matching hereafter. If $x$ is smaller than $L_{u}$,
but is still larger than a separation scale $x_{0}$, one can omit the
first term on the rhs of (\ref{smur}) and write the following general
solution of (\ref{smur}) in the intermediate domain, at $L_{u}>x\gg \sqrt{%
\kappa }$,

\begin{equation}
g\left( x,x^{\prime }\right) = 
{F_{1}+xF_{2},\,x<x^{\prime }, \atopwithdelims\{. D_{1}+xD_{2},\,x>x^{\prime }.}%
\label{inin}
\end{equation}
Here, the constants $F_{1},F_{2},D_{1},D_{2}$ should be fixed via inner
matching at $x=x^{\prime }$ and outer ones at $L_{u}$ and $x_{0}$. 
The value of $x_{0}$,
restricted by $\sqrt{\kappa }\ll x_{0}\ll \sqrt{\kappa }/s$, defines the
upper bound for the scales at which one cannot omit the first term of (\ref
{smur}) anymore. However, the point is that at $x<x_{0}<x^{\prime }$ an
actual dependance of $g$ on $x_{0}$ has disappeared 
and we should not locate $%
x_{0}$ explicitly to get the asymptotic behavior of $g(x,x^{\prime })$.
Indeed, integrating (\ref{smur}), we get 
\begin{equation}
\partial _{x}g\left( x\right) =s\int\limits_{0}^{x}\frac{g\left( z\right) dz%
}{S(x)+2\kappa },  \label{gint}
\end{equation}
where the argument $x^{\prime }$ is omitted and it is already taken into
account that $g(x)$ is even: $\partial _{x}g\left( x\right) |_{x=0}=0$. (\ref
{gint}) shows that the $x-$derivative changes sufficiently at the diffusive
scale, being nonetheless small by the absolute value there. It means that
the function $g\left( x\right) $ at $x\lesssim x_{0}$ is a constant in the
leading (zero) order in $s$. The subleading (first order in $s$) term
is then fixed by (\ref{gint}) taken at $x=x_{0}$. Alltogether it gives the
following expression for $g$ at $x<x^{\prime }$ with the desirable (for the
forthcoming matching) accuracy

\begin{equation}
g\approx g_{0}\left( 1+sx\int\limits_{0}^{x_{0}}\frac{dz}{S(x)+2\kappa }%
\right) \approx g_{0}\left( 1+s\frac{\pi x}{2\sqrt{2\kappa }}\right) ,
\label{proiz}
\end{equation}
where, evaluating the integral in (\ref{proiz}), we used that $\sqrt{\kappa }%
\ll x_{0}$. Matching (\ref{great},\ref{inin}) and (\ref{proiz}), we get 
\begin{equation}
\begin{array}{lll}
\mbox{at}\quad x=x^{\prime } & D_{1}+x^{\prime }D_{2}=F_{1}+x^{\prime }F_{2},
& F_{2}-D_{2}=1, \\ 
\mbox{at}\quad x=L_{u} & D_{1}+L_{u}D_{2}=A, & D_{2}=-\frac{\sqrt{s}}{L_{u}}%
A, \\ 
\mbox{at}\quad x=x_{0} & F_{1}+x_{0}F_{2}=g_{0}, & F_{2}=sg_{0}\frac{\pi }{%
2r_{d}}.
\end{array}
\label{jump}
\end{equation}
Keeping $s$ to be very small, while the ratios $s/\sqrt{\kappa }$ and $\sqrt{s%
}/L_{u}$ are finite, one derives from (\ref{jump})

\[
F_{1}=g_{0}=\frac{1+x^{\prime }\sqrt{s}/L_{u}}{s\pi /\left( 2r_{d}\right) +%
\sqrt{s}/L_{u}},\,\,\,F_{2}=\frac{sg_{0}\pi }{2\sqrt{2\kappa }}, 
\]

\begin{equation}
D_{1}=A=\frac{2\sqrt{2\kappa }+x^{\prime }s\pi }{s\pi +2\sqrt{2\kappa s}%
/L_{u}},\,\,\,D_{2}=-\frac{\sqrt{s}}{L_{u}}A.  \label{rshiv}
\end{equation}
It is worth noting that the explicit value of $x_{0}$ does not enter (\ref
{rshiv}), and the scale separation procedure is indeed justified.
Substituting (\ref{rshiv}) into (\ref{inin}) ($x^{\prime }$ $<L_{u}$), one
gets

\begin{equation}
\Omega _{s}\left( x\right) =\frac{1}{S\left( x\right) +2\kappa }\left[ \frac{%
\chi \left( x\right) }{s}-\frac{\chi \left( 0\right) }{s+\alpha \sqrt{s}}%
\right] ,\hspace{0.5cm}\alpha =\frac{2\sqrt{2\kappa }}{\pi L_{u}},
\label{otvs}
\end{equation}
and coming to the inverse Laplace-transform

\begin{equation}
\Omega \left( t,x\right) =\frac{1}{S\left( x\right) +2\kappa }\left[ \chi
\left( x\right) -\chi \left( 0\right) \frac{2}{\pi }\int\limits_{0}^{\infty
}dp\frac{\exp \left( -\alpha ^{2}p^{2}t\right) }{p^{2}+1}\right] .
\label{otvt}
\end{equation}
The expression 
(\ref{otvt}) states that, at the times less than ${\cal T}_{L_{u}}\sim \frac{%
L_{u}^{2}}{\kappa }$, the correlation function coincides with the
quasi-stationary limit given by (\ref{solution}). At the largest times, $%
\Omega \left( t,x\right) $ approaches the asymptotic value

\begin{equation}
\frac{1}{S\left( x\right) +2\kappa }\chi \left( x\right)  \label{as}
\end{equation}
diffusively, $\chi (x)-\Omega [S\left( x\right) +2\kappa ]=C\rightarrow \chi
\left( 0\right) \left( \alpha ^{2}\pi t\right) ^{-1/2}$.

\end{document}